\input mnrass.sty
\pageoffset{-2.5pc}{0pc}
 

\Autonumber  


\pagerange{000--000}
\pubyear{1996}
\volume{000}

\begintopmatter  

\title{A pulsational approach to the luminosity of Horizontal Branch stellar structures.}

\author{Renato De Santis \& Santi Cassisi}

\affiliation{Osservatorio Astronomico di Collurania, Via M. Maggini,
 I-64100, Teramo, Italy - E-Mail: desantis-cassisi@astrte.te.astro.it}

\shortauthor{R. De Santis \& S. Cassisi}
\shorttitle{The Horizontal Branch luminosity level.}


\abstract
\tx

We discuss an alternative approach to constrain the absolute bolometric 
luminosity of Zero Age Horizontal Branch (ZAHB) structures by using the observational pulsational properties 
of {\sl ab} type RR Lyrae stars and theoretical expectations concerning both the relation 
connecting the pulsational properties of these variables to their evolutionary ones, as luminosity, mass and 
effective temperature and, also the location in the H-R diagram for the fundamental pulsators instability
strip boundaries. Since the main goal of this work is to obtain an evaluation of the ZAHB bolometric
luminosity as much as possible independent on stellar evolution theory, we have minimized the use of
evolutionary prescriptions, being the only adopted evolutionary input the allowed mass range for fundamental
pulsators. Nevertheless, the effects on our final results related to the use of these evolutionary
prescriptions have been carefully checked.

In order to test the accuracy of the current framework, we have carefully investigated on the
effective temperature scale provided by De Santis (1996) and adopted in the present work. As a result, it has
been also found that the pulsational color - effective temperature scale fixed by the De Santis (1996) temperature
scale and the relation between intrinsic $(B-V)_0$ color and blue amplitude and metallicity as given by
Caputo \& De Santis (1992), appears fully consistent with both theoretical color transformations as given by
Castelli, Gratton \& Kurucz (1997a,b) and Buser \& Kurucz (1978) and, semiempirical one as provided by Green
(1988). On the contrary, it seems to exist a large discrepancy between this pulsational color-temperature
scale and the theoretical one based on the model atmospheres of Kurucz (1992). We have also analized the
effects on current analysis due to the adoption of a different effective temperature scale, namely the one
provided by Catelan, Sweigart \& Borissova (1998).

The reliability of the suggested method to obtain the ZAHB luminosity is shown by
applying it to a selected sample of globular clusters (GCs), whose heavy elements abundance covers almost 
all the complete GCs metallicity range. In order to verify the accuracy of the results obtained by using the
fundamental pulsators RR Lyrae stars, a quite similar analysis has been also performed by using both
observational and theoretical evidences for first overtone variables. 

The results obtained for the ZAHB bolometric luminosities in the adopted sample of clusters have been
critically analized and a comparison with evolutionary prescriptions on such quite important quantity 
as given by recent evolutionary computations has been also performed. The existence of evident mismatches
between current results and some evolutionary models has been verified and discussed. 

Finally, our investigation on the ZAHB luminosity levels has been extended to field variables, 
in order to check if it exists a {\sl real} difference in luminosity between cluster RR Lyrae stars 
and field ones as suggested by different authors through Hipparcos-based investigations. 
The comparison between the pulsational properties of field and cluster variables do not show the 
existence of any significant difference in their intrinsic luminosity, thus providing further support 
to the results obtained by Catelan (1998).


\maketitle  


\section{Introduction}
\tx 

The distance scale of low-mass, metal-poor stars plays an important role 
in stellar astrophysics, since the distance evaluations of old stellar 
systems such -as the globular clusters- are a fundamental step for 
providing accurate age determinations (see e.g. the detailed discussion 
by Renzini 1991)  and in turn for providing a firm lower limit to the age 
of the Universe.  
The traditional distance ladder for metal-poor populations is the luminosity 
of Horizontal Branch (HB) stars and in particular of RR Lyrae variables, 
which are currently considered the natural Population II 
{\em standard candles}. 

However, even though large observational and theoretical efforts have been 
devoted for estimating their \lq{true}\rq\ luminosity this quantity is still 
affected by large uncertainties. In particular, as recently suggested by 
Catelan (1998, and references therein) it seems that the luminosity of HB 
stars presents a \lq{dichotomy}\rq\ between a {\sl faint} and a {\sl bright}
distance scale. The former produces for the galactic GCs short distances and 
therefore large ages, whereas the latter long distances and small ages.
This problem has been widely discussed in the literature due to the release 
of the {\sl Hipparcos} database. In fact, the accurate measurement of stellar 
parallaxes should allow us to discriminate between the {\sl faint} and the 
{\sl bright} distance scale of RR Lyrae stars. 

Unfortunately, this important goal has not been accomplished, since recent
investigations on GCs based on Hipparcos parallaxes seem to support both 
the {\em bright} RR Lyrae distance scale (Gratton et al. 1997; 
Reid 1997, 1998; Chaboyer et al. 1998) and the {\em faint} distance scale 
(Pont et al. 1998). Further support to the the long RR Lyrae distance scale 
was brought out by recent analysis of different groups of variable stars 
such as Sx Phoenicis (Mc Namara 1997), Cepheids (Feast \& Catchpole 1997,
Madore \& Freedmann 1997) and Mira (van Leeuwen et al. 1997). 
Similar conclusions were also reached by new investigations based on the 
Red Giant Branch Tip (Salaris \& Cassisi 1998) and on the SN1987a ring 
(Panagia 1998; but see also Gould \& Uza 1998). 

This notwithstanding, accurate analysis of the Hipparcos parallaxes of 
field RR Lyrae stars support the faint RR Lyrae distance scale (Gratton 1998; 
Fernley et al. 1998; Tsujimoto, Miyamoto \& Yoshii 1998). From the above 
discussion emerges quite clearly that the problem of the luminosity of HB 
stars has not been properly settled yet. In fact, Gratton (1998) argued 
that the Hipparcos results could be explained if the luminosity of field 
and cluster RR Lyrae variables presents an intrinsic difference of the order 
of 0.2 mag. 
An accurate investigation of this problem has been recently performed by 
Catelan (1998), who showed that the pulsational properties of RR Lyrae in
the field and in GCs do not support the Gratton's conclusion concerning the 
existence of a dichotomy between the two different samples. 

The pulsational properties of RR Lyrae variables provide an unique 
opportunity for testing the prescriptions of both stellar evolution and 
stellar pulsation theories. In fact, fundamental constraints on the 
evolutionary properties of HB stars can be derived by adopting the pulsation 
relation i.e. the relation which supplies the pulsational period as a 
function of stellar mass, luminosity and effective temperature. 
The pulsation relation for RR Lyrae variables originally derived 
by van Albada \& Baker (1971) on the basis of linear, radiative, 
nonadiabatic models has been recently revised by Bono et al. (1997) 
and by Caputo, Marconi \& Santolamazza (1998a) by adopting full amplitude, 
nonlinear, convective models. Other interesting properties of RR Lyrae pulsation 
behavior have been brought out by Sandage, Katem \& Sandage (1981) who
suggested the existence of a tight correlations between temperature and 
pulsational amplitudes and by Caputo \& De Santis (1992, hereinafter CDS92) 
who supported the evidence of a clear correlation between period, blue 
amplitude and light-mass ratio. 

In this context it is worth mentioning that both periods and amplitudes 
can be measured with high accuracy, since they are affected neither by 
distance uncertainties nor by interstellar reddening evaluations. 
As a consequence, the use of these observables and the comparison between 
theory and observations can supply independent and useful constraints 
on the intrinsic luminosity of RR Lyrae variables. A similar approach 
was adopted by Castellani \& De Santis (1994) who showed that the 
absolute visual magnitude predicted by theoretical models are, within 
an accuracy of 0.1 mag., in satisfactory agreement with observed values. 

Quite recently, De Santis (1996, hereinafter DS96) provided an accurate 
revision of the RR Lyrae temperature scale adopted in previous 
investigations and suggested a slight change in the zero point of the 
magnitude scale for these variables. However, Caputo et al. (1998b) by 
investigating the pulsational and evolutionary properties of RR Lyrae 
variables in M5 found a sizable discrepancy between recent theoretical  
prescriptions and observations. According these authors, this discrepancy could be 
explained, without changes in the adopted evolutionary scenario, by slightly shifting 
the position in the H-R diagram of the instability strip toward higher effective 
temperatures.

The evaluation of HB luminosity based on stellar models has been recently 
reviewed by Cassisi et al. (1998a,b). In these papers, it has been thoroughly 
discussed the dependence of HB luminosity on the input physics adopted for 
computing stellar models. However, the current observational scenario 
does not supply sound constraints on HB magnitude and in turn on theoretical 
predictions based on different physical assumptions (De Boer, Tucholke \& 
Schmidt 1997; Cassisi et al. 1998a, 1998b). As a consequence, new and 
independent approaches are necessary for testing the accuracy of the 
theoretical scenario for low-mass helium burning stars and for assessing 
their intrinsic properties. 

The main goal of this investigation is to analyse the intrinsic luminosity 
of HB stars by adopting the pulsational characteristics of fundamental 
RR Lyrae variables (hereinafter $RR_{ab}$) in a sample of galactic GCs. 
It is worth stressing that for comparing theory and observations we 
did not transform into the observative plane theoretical observables, 
but we directly estimated the ZAHB bolometric magnitude.
As a consequence, this approach overcomes the uncertainties which affect 
the bolometric corrections based on static atmosphere models.
The theoretical framework and the method adopted for deriving the ZAHB 
bolometric magnitude are outlined in the \S 2. 

In \S 3, we briefly review our approach for estimating the effective 
temperatures of $RR_{ab}$ stars on the basis of their pulsational 
properties. In this section we also compare our results with the most 
recent theoretical color-temperature relations and discuss the evaluation
of the interstellar reddening for the selected clusters.
The results of the application of our method are presented in \S 4. 
Finally, observational data are compared with theoretical predictions 
and a critical analysis of the aftermaths of this investigation is 
presented. 

\section{The theoretical framework and the method.}
\tx

In this section, we discuss the theoretical framework and the procedure adopted for providing an
empirical evaluation of the ZAHB luminosity in a sample of galactic
globular clusters. 

Since we are interested in obtaining an independent measurement of the intrinsic luminosity of the ZAHB 
for providing firm constraints on the reliability of the current evolutionary scenario, we have reduced 
as much as possible the use of evolutionary models. 
In fact, our analysis relies only on the estimate of the stellar masses along 
the HB which produce fundamental pulsators.
However, it will be shown that our
results are largely unaffected by this approach. In every case, it will be accurately discussed the
uncertainty related to the use of these evolutionary evidences.

\figure{1}{d}{140mm}{\bf Figure 1. \rm Fraction of time of the global HB evolutionary 
lifetime spent at the various effective temperatures for several evolutionary models and for different
assumptions on the heavy elements abundance. The metallicity, the lower mass plotted and the mass step
between different tracks are labelled in each panel. The instability strip boundaries as provided by
Bono et al. (1997): fundamental red edge (solid line), fundamental blue edge (heavy solid line) 
and first overtone red edge (dashed line), are also shown. In the panel corresponding to
$Z=0.0002$, it has been also plotted the blue edge of the first overtone instability strip 
(heavy dashed line) (see text for more details).}
In the literature to account for both evolutionary and pulsational properties 
of RR Lyrae stars it has been generally adopted the pulsation relation 
provided by van Albada \& Baker (1971). However, recent theoretical 
investigations strongly support the evidence that sound estimates of both 
the pulsation period and the modal stability of RR Lyrae stars can only be 
obtained in a nonlinear regime (Bono \& Stellingwerf 1994). 
On the basis of an extensive grid of RR Lyrae full amplitude, nonlinear 
and time-dependent convective models Bono et al. (1997) 
have revised the pulsation relation by van Albada \& Baker. Therefore, 
the pulsation relation for fundamental pulsators, we adopt in the following, 
is:

$$\log{P}=11.627 + 0.823\cdot\log{L} - 0.582\cdot\log{M} - 3.506\log{T_e}\,\,\,1)$$

\noindent
where P is the fundamental period (days), $T_e$ is the effective temperature
and L and M are the luminosity and the stellar mass in solar units (see 
Bono et al. (1997) for more details). 

Since the Horizontal branch is not really {\sl horizontal},
we need to fix an effective temperature to which we will refer to when estimating the ZAHB luminosity. 
For this goal we decide to adopt an effective temperature equal to $\log{T_e}=3.85$ and, define 
as $L_{3.85}^{zahb}$ the ZAHB luminosity level at this effective temperature. Then we write the luminosity of each variable as: 

$$\log{L}=\log{L_{3.85}^{zahb}} + \Delta\log{L_{zahb}}$$

\noindent
where  $\Delta\log{L_{zahb}}$ is the difference in luminosity between the individual variable 
and the ZAHB at $\log{T_e}=3.85$. With simple algebric substitutions, equation 1) can be rewritten 
(see also Sandage 1981) as follows:

$$\log{P} - 0.823\cdot\Delta\log{L_{zahb}} = \log{P} + 0.33\cdot\Delta{M_{Bol}^{zahb}} =
\,\,\,\,\,\,\,
11.627 + 0.823\cdot\log{L_{3.85}^{zahb}} - 0.582\cdot\log{M} - 3.506\log{T_e} \,\,\, 2)$$

\noindent
where the symbols have their usual meaning, and the quantity 
$\log{P} + 0.33\cdot\Delta{M_{Bol}^{zahb}}$ corresponds to the fundamental reduced period (Sandage
1981). 

The method, we have developed for estimating the bolometric ZAHB magnitude, is based on the use
of the equation 2) in order to evaluate the relation between the fundamental reduced period and the
effective temperature, as a function of the variable mass (see below) and $L_{3.85}^{zahb}$, and
on the comparison between this relation and the observational prescriptions on these same quantities 
as provided by a sample of $RR_{ab}$ variables in selected GCs. In section IV, this procedure will be
outlined in more detail and it will be shown that this method allows us to obtain a straightforward
estimate of the ZAHB luminosity (i.e. $L_{3.85}^{zahb}$).

In order to compare theory and observations, we need to transform the equation 2) into the observative 
plane, so we adopt:

$$\Delta{M_{Bol}^{zahb}}=\Delta{V^{zahb}} + \Delta{BC}$$

\noindent
where we have defined the difference in bolometric correction between the individual variable and the
ZAHB at $\log{T_e}=3.85$ as $\Delta{BC}=BC_{\log{T_e}} - BC_{3.85}$.
Obviously the bolometric correction in the previous relation has to be 
evaluated by adopting the surface gravity and effective temperature values 
of the variable star and of the fictious star located on the ZAHB at $\log{T_e}=3.85$. 
However, it is worth noting that, in the CMD region populated by RR Lyrae 
stars, the allowed gravity range is quite narrow, as it can be easily tested by using any set
of evolutionary models available in the literature and a theoretical evaluations of the 
instability strip boundaries.
However, in order to validate our assumption - i.e. to neglect the gravity 
changes within the instability strip - we performed a test by adopting 
several sets of bolometric corrections (see below).
Interestingly enough we found that the bolometric correction is sensitive 
to surface gravity values but the quantity $\Delta{BC}=BC_{\log{T_e}} - BC_{3.85}$ 
presents a negligible dependence on gravity in the range of effective 
temperatures and luminosities typical of $RR_{ab}$ stars.
\figure{2}{S}{80mm}{\bf Figure 2. \rm Comparisons in the $(B-V)-A_B$ plane between the intrinsic
colors of M3 RR Lyrae stars as obtained by the CDS92 relation (solid line)
and the one obtained by using the $T_{eff}$ scale of De Santis (1996) and theoretical (panel a: CGK97
- panel c: BK78 - panel d: K92) or semi-empirical (panel b: Yale) color - temperature relations.}

Finally we have adopted the following relation for $\Delta{BC}$ as a function of the effective
temperature, derived by adopting the Buser \& Kurucz (1978) model atmospheres:

$$\Delta{BC}= -5.252\cdot{\log}^2{T_e} + 41.636\cdot\log{T_e} - 82.454 \,\,\,\,\,\,\,\,\,\,3)$$

\noindent
The standard deviation of this relation is $\sigma=\pm0.004$, while the correlation coefficient $r\approx1.0$ 
and it is correct in the ranges: $6000\le{T_e(K)}\le7500$, $2.5\le\log{g}\le3.0$ and metallicity $0.0002\le{Z}\le0.002$. 
It is important to notice that the use of a different BC scale does not substantially change the dependence of $\Delta{BC}$ on temperature
as given by equation 3).

The next topic, we wish to address, is the one corresponding to the method adopted to evaluate the
visual magnitude of the ZAHB from the observations of RR Lyrae variables and non-variable HB stars
in globular clusters. 

In order to estimate the ZAHB visual magnitude, we have adopted the same operative approach described 
by Sandage (1990) by taking into account the {\sl lower envelope} of the HB star distribution. 
It could be raised the question that in the
previous relations we have adopted the visual magnitude of the ZAHB at $\log{T_e}=3.85$ while 
now we adopt the \lq{\sl generic}\rq\ lower envelope of the HB star distributions. 
Even though these definitions seem to be substantially different, any 
test performed by adopting the grids of evolutionary models 
(Dorman, Rood \& O'Connell 1993; Caloi, D'Antona \& Mazzitelli 1997; 
Cassisi et al. 1998ab) and the sets  of color-temperature relations and 
of bolometric correction scales available in the literature, 
discloses that within the instability strip the dependence of $M_V(ZAHB)$ 
on the effective temperature is quite negligible. In fact, we found that  
$M_V(ZAHB)$ changes by only 0.006 mag when moving from $T_e\approx6700$K 
to $\approx7100$K.

As a relevant point, we note that our method is not affected by uncertainties
on stellar photometry  such as the zero point and the calibration procedure, since we use the apparent
visual magnitude of the ZAHB as a reference magnitude from which to measure the mean visual
magnitude of each variable. In fact our approach relies only on the difference in the V magnitude between each
{\sl ab} RR Lyrae star and the ZAHB at approximately $\log{T_e}=3.85$.

As we have underlined at the beginning of this paper, the main goal of our work is to obtain an
evaluation of the luminosity level of the HB stars in GCs as much as possible independent on the
theoretical evolutionary scenario. As it can be easily recognized in all previous relations which
represent our working framework, do not appear quantities related to evolutionary theory but the
mass of the variable.
\figure{3}{S}{80mm}{\bf Figure 3. \rm As figure 2, but for the {\sl ab} RR Lyrae stars in M15.}

It is well known that the mass of an RR Lyrae variable can not be directly evaluated unless the
star is a double pulsators and, also in this case such evaluation is based on theoretical
pulsational models (Kovacs, Buchler \& Marom 1991). For long time, it existed a large discrepancy between the  pulsational 
and evolutionary determinations of double mode RR Lyrae stars mass, but recently the introduction of new
radiative opacity evaluations from the Livermore Laboratory (OPAL opacity, Rogers \& Iglesias 1992)
has allowed to make insignificant this discrepancy. 

In the present work, we adopt evolutionary determinations of the pulsator masses as provided by our
own stellar computations (Bono, Cassisi \& Castellani 1998). Since as it is evident, we can not assign a
mass to the whole sample of variables in a GC, we use only an evaluation on the {\sl lower} and {\sl upper} mass
($M_{min}^{RR}$ and $M_{max}^{RR}$ respectively) producing {\sl ab} RR Lyrae stars. Due to the
dependence on the metallicity of these parameters, we have analized their variations when changing
the heavy elements abundance in a quite large range, namely $0.0001\le{Z}\le0.003$.

For each fixed metallicity, we have adopted as $M_{min}^{RR}$ and $M_{max}^{RR}$ the lower and
larger mass respectively which spend in the instability strip as fundamental pulsators at least the 
5\% of their central He burning evolutionary lifetime. 
In order to adopt a homogeneous pulsational context the boundaries of the instability strip were
fixed according to the prescriptions by Bono et al. (1997) i.e., the same set of RR Lyrae models
adopted for the pulsational relation (equation 1).

In figure 1, it is shown for several metallicities, the fraction of the total HB evolutionary lifetime spent 
by HB stars at the various effective temperatures. Data plotted in this figure clearly prompt that the 
assumption to take into account only models which spend in the fundamental 
instability strip a time longer or equal to 5\% of their HB evolutionary 
lifetime does not affect our final results.

It is worth emphasizing that we also assume, according to Bono, Caputo \& Stellingwerf (1994), 
that inside the \lq{OR}\rq\ region of the instability strip - i.e. the region where both 
fundamental and first overtone pulsators attain a stable limit cycle - the pulsation mode is 
governed by the previous evolutionary history of the variable (hysteresis mechanism).

At the same time, we note that the dependence of periods on pulsators
masses (see equation 1) is very weak, and indeed an uncertainty of $\approx0.01M_\odot$ on the mass 
implies an indeterminations on $log{L_{3.85}^{zahb}}$ of the order of 0.003, when the value of the
all other quantities is fixed. 
In addition, one has to take into account that to determine the HB luminosity level in a cluster, we
adopt, when possible, a quite rich sample on RR Lyrae stars and this occurrence contributes in reducing the final
error.

\section{The effective temperature scale.}
\tx

The analytical relation which supplies the effective temperature as a function of period and
blue amplitude ($A_B$), we adopt in this investigation, is the one provided by DS96 
which corresponds to an updated version of the relation given by Castellani \& De Santis (1994):

$$\log{T_e}= -0.1094\cdot\log{P} + 0.0134\cdot{A_B} + 3.770 \,\,\,\,\,\,\,\,\,\,\,\,\,\, 4)$$

\noindent
with a probable error equal to $\pm0.003$. 
This relation has been derived by relying on the RR Lyrae database of
Lub (1977) and, by correcting its zero point in order to achieve agreement with the
Baade-Wesselink data for field variables (Carney, Storm \& Jones 1992, hereinafter CSJ92) 
(see DS96 for more details on this point).

The accuracy of this or a similar relation for deriving the effective temperature of RR Lyrae 
stars has been recently questioned by Walker (1998) due to the large number of parameters that are involved 
and to the small number of calibrating stars. However, this is not the case for equation 4), since it has been
obtained by using a quite large sample (about 70 objects) of RR Lyrae stars.
\figure{4}{S}{80mm}{\bf Figure 4. \rm As figure 2, but for the {\sl ab} RR Lyrae stars in M68.}

Both the reliability and accuracy of this effective temperature scale has already been extensively discussed by
Castellani \& De Santis (1994) and, due to the revision, further emphasized by DS96.
Therefore, this topic will not be again reviewed here and we address the interested reader to 
the quoted papers. However, it is worth remembering that CDS92, through an 
accurate analysis of the Lub's (1977) data for field RR
Lyrae stars with well-determined reddening, have clearly shown the existence of a tight
correlation between the intrinsic color of the variable and its blue amplitude and metal
abundance:

$$(B-V)_0= -0.0775\cdot{A_B} + 0.005\cdot[Fe/H] + 0.434 \,\,\,\,\,\,\,\,\,\,\,\, 5)$$

\noindent
with a standard deviation equal to $\sigma=\pm0.015$ mag. 

Obviously, the coupling of equations 4) and 5) provide a color - temperature scale, so it could
be interesting to check the consistency between these two equations by comparing it with
the color - temperature relation provided by theoretical model atmospheres.

Being aware of the problems still existing with theoretical model atmospheres
(Castelli 1998, private communication), we have tried to perform this comparison by using as
many as possible independent sets of transformations: Buser \& Kurucz 1978 (hereinafter BK78), 
Kurucz 1992 (K92) and Castelli, Gratton \& Kurucz 1997a,b (CGK97).
In order to provide a comprehensive analysis on the accuracy of color - temperature scales, we also 
adopt here the Yale semi-empirical transformations (Green 1988). 
These transformations are an empirical UBVRI recalibration of Vandenberg \& Bell (1985) and 
Kurucz (1979) synthetic colors and $BC_{V}$, based on various observational 
constraints.

To perform the check we have used the observational data for {\sl ab} RR Lyrae stars in 3
globular clusters, namely M3, M15 (Sandage 1990) and M68 (Walker 1994). The procedure is quite
simple: i) we derive the effective temperatures of the variable stars by using their period and
B amplitude (equation 4); ii) their intrinsic $(B-V)$ colors are then estimated by adopting an average 
value for the gravity i.e. $\log{g}=2.75$ and by using a color - temperature relation (see below) for the suited
metal abundance ([Fe/H]=-1.3 for M3 and [Fe/H]=-2.0 for M15 and M68).
These color estimations are then compared with the ones provided by equation 5) in the $(B-V)_0
- A_B$ plane, as shown in figures 2) to 4).

It is worth noting the fine agreement achieved when comparing the data obtained by adopting the
relation provided by CDS92 with the colors obtained by using both the theoretical color-temperature
scales by BK78 and CGK97 and also with the Yale semi-empirical one. The average difference is equal
to 0.005 mag in the worst case (panel a in fig. 3). On the contrary, it is important to notice the
large discrepancy obtained when using the theoretical color-$T_{eff}$ relation from K92.
In this case the average difference is of the order of $\approx0.04$ mag. 
These results led some support to the following conclusions:
\medskip
\noindent
i) {the color ($B-V$) - effective temperature relation provided by equations 4) and 5), 
appears fully consistent with both the theoretical scales based on the model atmospheres by 
BK78 and CGK97 and the semi-empirical one by Green (1988). This occurrence suggests that the Lub's data
for field RR Lyrae stars and, therefore, his interstellar reddening scale, are quite consistent with the
previous quoted color - $T_{eff}$ scales;}
\smallskip
\noindent
ii) {it seems to exist a large discrepancy between the color -$T_{eff}$ scale based on model
atmospheres from K92 and our \lq{pulsational}\rq\ scale based on equations 4) and 5). This result is
a clear evidence that the reddening scale of Lub (1977) on which rely both equation 4) and 5) is not
consistent with the color-temperature scale of K92. It is evident that such occurrence could be
explained as a drawback of the K92 model atmospheres or alternatively as due to a problem in the
Lub's reddening scale.}
\medskip
\noindent
In this context, it is worth noticing that recently several
accurate theoretical and observational investigations have provided clear indications about the presence
of significant problems in comparing theory and observations when using the color-temperature relation as
given by Kurucz (1992). In particular, in a detailed analysis of the pulsational properties 
of RR Lyrae stars in GCs, Bono et al. (1997) have clearly shown that the use of the K92 color-temperature 
scale produces, for each given amplitude, effective temperatures about 300K higher than predicted by the
theory. When considering the fine agreement that the quoted authors, achieved as far as it concerns the
Bailey (blue amplitude {\it versus} period) diagram, one is strongly forced to consider this result as a
clear evidence for some problems in the K92 color-temperature relation. This question has been also
addressed by Silbermann \& Smith (1995) by comparing the $(V-R)$ color of a sample of RR Lyrae stars
with theoretical expectations. In addition, it is important to remember that an accurate analysis on the
possible drawbacks in model atmosphere computations and, more in detail, on the K92 calculations, have
been provided by CGK97 and by Castelli (1998, private communications).

Taking into account such evidences, we suggest that our result is a further indication
against the use of the K92 color-temperature scale, at least when working on RR Lyrae variables.

It is also worth noticing that the satisfactory agreement achieved between the results provided by
equations 4) and 5) and the Yale, BK78 and CGK97 color-temperature scales, can be regarded as a positive
check on the consistency and reliability of the pulsational effective temperature scale adopted in the
present work.
\figure{5}{S}{80mm}{\bf Figure 5. \rm Comparison in the $(B-V) - A_B$ plane between static colors
as obtained by using the K92 (open circles) and CGK97 (full circles) color - temperature scales,
corresponding to the pulsational models of Bono et al. (1997) for different assumptions on the luminosity
level but for the same pulsator mass $0.65M_\odot$ and metallicity Z=0.001. The CDS92 relation is also
displayed.}

Up to now we discussed the accuracy of the color - temperature scale defined by equations 4) and 5), now we
briefly address a question concerning the use of mean colors and effective temperatures.
The relation between effective temperature, period and blue amplitude
(equation 4) has been obtained by adopting for each variable in the Lub's sample the mean $(B-V)$ color, i.e.
the time average along a full pulsational cycle of the color curve (see DS96 for more details on this point). 
At the same time, 
the effective temperature adopted by Bono et al. (1997) for deriving 
the pulsational equation 1) is the \lq{\sl static}\rq\ equivalent temperature.
As a consequence, it could be claimed that we are dealing with two \lq{different}\rq\ and, un-consistent
 effective temperature definitions. However, DS96 has already shown
the quite fine agreement between the effective temperature evaluations provided by equation 4) and
the static temperatures from CSJ92. 

Concerning this point, it is also worth noticing that the comparison (see for instance Bono, Caputo \& 
Stellingwerf 1995) between static colors and mean (along the pulsational cycle) ones can not be adopted 
at all to check the consistency between static and mean effective temperatures. This occurrence is due to
the evidence that the obtained results are strongly dependent on the adopted color - temperature relation.
This effect is shown in figure 5, where we have plotted the static colors corresponding to the pulsational
models provided by Bono et al. (1997), as obtained by adopting two different color -temperature scales,
namely the K92 and the CGK97 one. For the aim of comparison, it has been also displayed the CDS92
relation. As a result, one can easily notice the dramatic effect related to the use of different 
color - $T_{eff}$ relations.
When considering these or similar results, one should use cautiously every theoretical relation,
allowing to derive static colors from mean color evaluations.

Before closing this section, we wish to adopt the previously discussed relation between intrinsic color
and pulsational properties of the variables in order to evaluate the interstellar reddening for a sample
of clusters. In fact, even though our approach for estimating the HB luminosity does not depend on the
reddening, the color - temperature scale, we derived, can be easily adopted for providing independent
estimates of this parameter.

The sample of GCs taken into account, is the same adopted for the further investigation on the HB
luminosity level and, it will be discussed in the next section.
\figure{6}{S}{60mm}{\bf Figure 6. \rm Comparison in the $(B-V) - A_B$ diagram between the observed colors
of a sample of RR Lyrae stars in the cluster M3 and the reddening free colors (dashed line) as given by
CDS92 relation. The heavy solid line corresponds to the CDS92 relation vertically shifted in order to
match the observational data.}

The procedure adopted to derive the reddening is quite simple: we compare in the $(B-V) - A_B$ plane
the observational data for the sample of {\sl ab} RR Lyrae stars in each cluster with the reddening free
colors provided by the equation 5) by using the observational blue amplitudes and the cluster metallicity
$[Fe/H]$ as given by Carretta \& Gratton (1997, hereinafter CG97). The amount of color shift which is need 
in order to achieve a satisfactory agreement between observations and the CDS92 relation, provides a 
fine evaluation of the cluster reddening.
In figures 6) to 8), we show the comparison between the observed color of the RR Lyrae stars and the
dereddened color as provided by the CDS92 relation for the clusters: M3, M15 and M68, respectively.

\figure{7}{S}{60mm}{\bf Figure 7. \rm As figure 6) but for the cluster M15.}

The reddening evaluations obtained with this method and their maximum errors, are listed for all clusters 
in our database in Table 1) together with other relevant cluster parameters (see below).

In passing, we note that our evaluation of the M68 reddening is in fine agreement with the estimate recently provided
by Gratton et al. (1997) on the basis of their accurate Str\"omgren photometric analysis of field stars projected on
the sky near a selected sample of GCs.
\figure{8}{S}{60mm}{\bf Figure 8. \rm As figure 6) but for the cluster M68.}
\table{1}{S}{\bf Table 1. \rm Selected data for the sample of galactic
globular clusters.} 
{\halign{%
\rm#\hfil & \hskip3pt\hfil\rm#\hfil & \hskip2pt\hfil\rm#\hfil & \hskip4pt\hfil\rm#\hfil & \hskip4pt\hfil\rm\hfil#\cr
Cluster& [Fe/H] & $V_{zahb}$ & $\log{L_{3.85}^{zahb}}$ & E(B-V) \cr 
\noalign{\vskip 10pt}
NGC1851 & -1.08 & $16.13\pm0.025$ & $1.645\pm0.010$ & $0.045\pm0.016$ \cr 
NGC4590 (M68) & -1.99 & $15.70\pm0.02$  & $1.70\pm0.02$ & $0.043\pm0.014$   \cr
NGC5272 (M3) & -1.34 & $15.73\pm0.02$  & $1.645\pm0.010$ & $0.012\pm0.010$ \cr 
NGC6171 & -0.87 & $15.85\pm0.10$  & $1.54\pm0.04$ & $0.383\pm0.023$   \cr 
NGC6362 & -0.96 & $15.33\pm0.03$  & $1.615\pm0.015$ & $0.070\pm0.013$ \cr
NGC6981 & -1.30 & $17.07\pm0.03$  & $1.635\pm0.010$ & $0.070\pm0.015$  \cr 
NGC7078 (M15) & -2.12 & $15.92\pm0.05$  & $1.68\pm0.02$ & $0.074\pm0.014$   \cr}}

\section{\bf The ZAHB luminosity level.}
\tx

\subsection{Fundamental pulsators.}
\tx
In order to apply the method for estimating the ZAHB luminosity we selected 
a sample of galactic GCs which is characterized by homogeneous photometric 
observations for both non-variable HB stars and RR Lyrae variables and 
that covers a wide range of metal contents. 
On the basis of these requirements we selected the following clusters: M3, M15, M68, NGC6171,
NGC1851, NGC6362 and NGC6981. The data for both variable and non-variable stars in M3, M15,
NGC6171 and NGC6981 have been collected from the work of Sandage (1990); for M68, NGC1851 and NGC6362
we have used the data from Walker (1994), Walker (1998) and Brocato et al. (1998), respectively.
Unfortunately for some of these clusters only the photographic photometry of static and variable 
HB stars is available in the literature. 
However, as already mentioned, the lack of accurate CCD data is not  
a strong limit to the application of our method. In fact, our analysis 
relies on the difference in magnitude between the RR Lyrae stars and 
the ZAHB. This means that the ZAHB luminosities we derive are not affected 
by any uncertainty on the {\sl true} zero point of the photometry.
\figure{9}{D}{150mm}{\bf Figure 9. \rm The distribution in C-M diagram of the HB stars (open circle:
non-variable stars; full circle: RR Lyrae stars) for all selected clusters. In each panel, the adopted visual
magnitude of the ZAHB with the associated error is shown.}

For each cluster in our sample in order to apply the method outlined in section II, we have to 
preliminary estimate the ZAHB visual magnitude, and also the expected range of fundamental pulsator mass.
In the following, we discuss for each single cluster the values adopted for these important parameters:
\medskip
\noindent
{\bf NGC1851}: for this cluster, a metallicity evaluation is not present in the CG97 compilation 
so we have adopted the estimate provided by Zinn \& West (1984). However, in order to be consistent with the CG97
metallicity scale adopted for all other clusters, we have taken into account the relation relating the Zinn \& West
(1984) scale with the CG97 one (the equation 7 in CG97), so finally we have adopted $[Fe/H]=-1.08$, 
which corresponds to $Z\approx0.002$. This occurrence allow us 
by using theoretical stellar models of suitable metallicity to obtain a mass
interval for {\sl ab} RR Lyrae stars in this cluster equal to: $0.61\le{M_{RRab}/M_\odot}\le0.67$.
Concerning the visual magnitude of the ZAHB, it has been estimated by
adopting the lower envelope of the observed distribution.
In figure 9, we have plotted for each cluster in our sample the observed distribution of stars on the HB
and marked the adopted visual magnitude of the ZAHB (see also Table 1).
By adopting this approach we estimate for the visual magnitude of the ZAHB ($V_{zahb}$) for the cluster NGC1851,
a value equal to $16.13\pm0.025$ mag.
\medskip
\noindent
{\bf M68}: in this case the metallicity measurement provided by CG97 is equal to $[Fe/H]\approx-1.99$, 
i.e. $Z\approx0.0002$. From data in figure 1), this mass range for
{\sl ab} RR Lyrae stars has been evaluated: $0.70\le{M_{RRab}/M_\odot}\le0.80$. 
By analizing the photometric data from Walker (1994), we estimate a value for the ZAHB visual magnitude 
equal to $15.70\pm0.02$ mag.
\medskip
\noindent
{\bf M3}: the heavy element abundance of this cluster according to the CG97 metallicity scale is equal to
$[Fe/H]=-1.34$ which corresponds to about $Z\approx0.001$. Therefore, by accounting for the theoretical
evidences shown in figure 1), we estimate that the most suitable fundamental pulsator mass range is:
$0.64\le{M_{RRab}/M_\odot}\le0.70$. 
The adopted value for $V_{zahb}$ is $15.73\pm0.02$ mag.
\medskip
\noindent
{\bf NGC6171}: by adopting $[Fe/H]=-0.87$ from the stellar models for $Z=0.003$, one obtains a
mass range for the fundamental mode variables equal to: $0.60\le{M_{RRab}/M_\odot}\le0.635$. From the data
plotted in figure 9), the most suitable estimate for $V_{zahb}$ is of about $15.85\pm0.10$ mag.
\medskip
\noindent
{\bf NGC6362}: for this cluster CG97 report a metallicity value equal to -0.96 dex,
which corresponds to $Z\approx0.002$. Therefore we adopt for this cluster a RR Lyrae mass range equal to:
$0.61\le{M_{RRab}/M_\odot}\le0.66$. As far as it concerns the visual magnitude of the ZAHB from the observed 
distribution of HB stars we derive a value of $15.33\pm0.03$ mag.
\medskip
\noindent
{\bf NGC6981}: since the metallicity of this cluster is equal to $\approx-1.30$, we adopt the same mass range
adopted for M3 and, by analizing the observed distribution of HB stars in the CM diagram we obtain a value
for $V_{zahb}$ of about $17.07\pm0.03$ mag.
\medskip
\noindent
{\bf M15}: since the metallicity of this cluster is quite similar to the M68 metallicity,
the same mass range for the fundamental pulsators has been adopted. From the data in figure 9), we estimate 
a value for $V_{zahb}$ equal to $15.92\pm0.05$ mag.

\medskip
\noindent
In section II, we have already
defined the fundamental reduced period: $\log{P_{red}}=\log{P} + 0.33\cdot\Delta{M_{Bol}(ZAHB)}$. Now, we
compare the observational data for the {\sl ab} RR Lyrae variables in each individual cluster with the 
prescription provided from equation 2) in the $\log{P_{red}} - \log{T_e}$ diagram. Since these theoretical 
expectations depend on the bolometric magnitude of the ZAHB at $\log{T_e}=3.85$ (according to our definition),
this comparison supplies a straightforward evaluation of the ZAHB luminosity. 
More in detail, the procedure adopted for estimating this quantity is the following: 
\smallskip
\item{$\bullet$}
{since equation 2) depends on the mass of the pulsators by adopting both the lower 
and upper limit of the $RR_{ab}$ mass range, we obtain two different solutions for the behavior of the 
reduced period as a function of the effective temperature, both depending on the ZAHB luminosity level;}
\item{$\bullet$} 
{then the ZAHB luminosity is estimated by properly fitting the lower and upper
boundaries of the RR Lyrae distribution in the reduced period - temperature plane. This approach is 
outlined in figure 10.} 
\smallskip
\noindent
\figure{10}{D}{140mm}{\bf Figure 10. \rm Comparison in the fundamental reduced period - temperature diagram
between the observational data for RR Lyrae stars in the selected GCs and the prescriptions of the
pulsational theory as given by equation 2 (see text for more details). In each panel, the two lines 
correspond to the evaluations provided by the equation 2, when adopting alternatively the upper (solid
line) or the lower limit (dashed line) for the allowed mass range for fundamental pulsators.}
The results listed in Table 1) support the following conclusions:
\smallskip
\noindent
{\sl i)} the bolometric ZAHB luminosity decreases significantly: $\Delta\log{L_{3.85}^{zahb}}\approx0.14$, when
increasing the metallicity from the most metal-poor cluster (M15) to the more metal-rich one  (NGC6171) 
in our sample;
\smallskip
\noindent
{\sl ii)} the ZAHB luminosity levels evaluated for the subsample of metal-poor clusters (M15 and M68) appear
in fine agreement. The same outcome applies for intermediate metallicity clusters  such as NGC6981 and
M3. This result can be regarded as an evidence of the reliability of our approach.
\smallskip
\noindent
{\sl iii)} the method, we developed, allows to estimate the bolometric ZAHB luminosity
at a fixed effective temperature with an high accuracy , and indeed in the worst case the uncertainty on
$\log{L_{3.85}^{zahb}}$ is of the order of 0.04; 
\medskip
\noindent

On the basis of the ZAHB luminosities we estimated for these clusters, 
we can transform the observational data into the theoretical plane, 
thus suppling an independent approach to compare theory and observations. 
We perform this comparison in the last section.

\subsection{First overtone pulsators.}
\tx

The evaluations of the ZAHB luminosity discussed in the previous subsection 
are based on $RR_{ab}$ variables. In order to provide an independent test
on the accuracy of the adopted method, we undertake a similar 
investigation but by adopting first overtone variables ($RR_c$).

Bono et al. (1997) have provided a relation, quite similar to the equation 1), but suitable for first
overtone pulsators:
$$\log{P}=10.789 + 0.800\cdot\log{L} - 0.594\cdot\log{M} - 3.309\cdot\log{T_e}\,\,6)$$

\noindent
where the symbols have their usual meaning and the units are the same as in equation 1).
By following the same definitions adopted in section II, one obtains:

$$\log{P} + 0.32\cdot\Delta{M_{Bol}^{zahb}} = 
10.789 + 0.800\cdot\log{L_{3.85}^{zahb}} + \,\,\,\,\,\,\,\,\,\,\,\,\,\,\,\,- 0.594\cdot\log{M} 
- 3.309\log{T_e} \,\,\,\,\,\,\,\,\,\,\,\,\,\,\,\,\,\,\,\,\,\,\,\,\,\,\,\,\,\,\,\,\,\,\,\,\,\,\,\,\,\,\, 7)$$

\noindent
Since the relation 4) has been obtained by adopting a sample of field $RR_{ab}$ Lyrae stars, it can not be
used for first overtone pulsators. So in order to estimate the effective temperature of $RR_c$ Lyrae
stars, we have adopted a different approach:
\smallskip
\noindent
{\sl i)} by using the reddening evaluation obtained by using the $RR_{ab}$ Lyrae stars (see section III and
data in Table 1), we have derived the {\sl true} $(B-V)$ color of the sample of first overtone pulsators 
in the selected clusters;
\smallskip
\noindent
{\sl ii)} from the $(B-V)_0$ color the effective temperature of the variable has been estimated by using
a theoretical color-temperature relation (as given, for instance, by CGK97 or by BK78 or Yale), since 
in section III we have shown the consistency between  these temperature scale and the pulsational one
corresponding to equations 4) and 5).
\medskip
\noindent
This method has been applied to the variables in two clusters, namely, M15 and M68. In the case of M15 we
have taken into account 26 $RR_c$ Lyrae stars (Sandage 1990), while for M68 we have excluded from
our analysis all variables affected by Blazko effect (see Walker 1994 for more details) and then we have
recovered only 15 variables.

As far as it concerns the mass range for first overtone pulsators, from data in figure 1) it has been obtained a
mass interval: $0.67\le{M_{RRc}/M_\odot}\le0.75$, for both clusters (their metallicity is quite similar).
The result of this investigation is shown for both clusters in figure 11.

\figure{11}{S}{90mm}{\bf Figure 11. \rm As figure 10, but for first overtone pulsators in M15 and M68.}

We have obtained a luminosity level of the ZAHB equal to $\log{L_{3.85}^{zahb}}=1.67$ and 1.69 for M15 and
M68 respectively, and then in fine agreement, within the associated uncertainties, with the previous
evaluations based on the pulsational properties of $RR_{ab}$ Lyrae stars. This occurrence can be
clearly considered an additional support to the reliability of our working framework and of the accuracy of the
adopted method.

\section{\bf Discussion and conclusions}
\tx

In the previous sections, it has been shown that the method, we developed, to derive 
from the pulsational properties of RR Lyrae stars in GCs, an estimate of the bolometric magnitude of 
the ZAHB at $\log{T_e}=3.85$  allow us to evaluate this important quantity with an high accuracy: the
average uncertainty on $\log{L_{3.85}^{zahb}}$ is equal to $\approx0.02$. Therefore, till now it can be
considered one of the most reliable approaches to constrain the HB luminosity level in GCs.

Nevertheless, it is important to provide a deeper insight on this method by showing the
main possible error sources in order to better quantify its accuracy. The most important uncertainties
can derive from: the evaluation of the visual magnitude of the ZAHB in GCs, the estimate of the suitable
mass range for fundamental pulsators, the adopted relation between pulsational period and the evolutionary
properties of the variable, the metallicity of the clusters and the adopted RR Lyrae
temperature scale. On this concern, it is quite easy to verify the following indications:
\medskip
\noindent
{\sl a)} an uncertainty on $V_{zahb}$ of about 0.025 mag produces an error on $\log{L_{3.85}^{zahb}}$
of about 0.01;
\smallskip\noindent
{\sl b)} as already discussed in section II, a shift of the whole mass range suitable for 
fundamental pulsators of $\Delta{M}=0.01M_\odot$ gives $\Delta\log{L_{3.85}^{zahb}}\approx0.004$. 
It is worth remembering that the effective temperature of stellar models, which is important in order 
to estimate the time spent inside the instability strip by each model, is usually strongly affected by 
all the physical inputs (as low temperature opacities and superadiabatic region treatment) which determine the stellar 
outer layers structure. On this concern, we wish to emphasize that all evolutionary computations adopted in
our analysis have been performed by using the most updated stellar physics as far as concerns stellar matter
opacity (see Bono et al. 1998 for more details), while for the treatment of the
superadiabatic layers it has been adopted the mixing length calibration provided by Salaris \& Cassisi
(1996). Nevertheless, it is important to verify if the estimations of fundamental pulsators mass ranges
($\Delta{M_{RRab}}$) at the various metallicities, obtained by using our own evolutionary computations are 
consistent with the results provided by stellar evolutionary computations on literature. More in detail, we
have compared our results with the ones obtained when using the Castellani, Chieffi \& Pulone (1991) 
models or the Caloi, D'Antona \& Mazzitelli (1997) ones. The comparison with the HB models of Castellani et al. (1991) has shown the existence of
a quite good agreement, for instance at Z=0.001, we obtain a mass range equal to 
$0.64\le{M_{RRab}/M_\odot}\le0.70$ which has to be compared with the range
$0.64\le{M_{RRab}/M_\odot}\le0.69$, one obtains when using the Castellani et al. (1991) models. As far as it
concerns the comparison with the HB models of Caloi et al. (1997), in the limit of the too coarse grid of models
adopted by these authors, a satisfactory agreement is obtained for all the metallicities for which is
possible to perform the comparison;
\smallskip\noindent
{\sl c)} as already discussed, the relation connecting the pulsational periods to
stellar masses, luminosities and effective temperatures as given by Bono et al. (1997), represents a
significant improvement in comparison with the original relation provided by van Albada \& Baker (1971).
However, it could be interesting to verify the effect on the present analysis due to the use of the
van Albada \& Baker's relation. Bono et al. (1997) have
already realized a comparison between their own relation and the van Albada \& Baker's one: due to the use
of more reliable nonlinear pulsational models they find periods smaller than linear periods (see their figure
6). By taking into account this occurrence, one can easily verify that the use of the van Albada \& Baker's
relation in order to constraint the ZAHB luminosity level in GCs has the effect to produce an average
reduction on the evaluation of $\log{L_{3.85}^{zahb}}$ of about $0.015-0.02$. It is also worth noting that 
recently Caputo, Marconi \& Santolamassa (1998) have revised the original Bono et al. (1997) relation by including 
the dependence on the heavy elements abundance. We have also investigated if the use of this revised relation
could significantly affect our results: it has been obtained that for the metal-poor clusters there are no effects
on the evaluation of the ZAHB luminosity and, that for metal-rich GCs the change on $\log{L_{3.85}^{zahb}}$ estimate
is, in the worst case, of the order of $\approx-0.01$;
\smallskip\noindent
{\sl d)} concerning the metallicity of the cluster, it is worth noticing that the only point where the
heavy elements abundance plays a role is related to the choice of the most suitable mass range for RR
Lyrae stars. However, if one takes into account the discussion at the point {\sl b)}, the
evidences shown in figure 1) and also that the mean uncertainty in the CG97 metallicity scale 
($\approx0.10-0.15 dex$), it is easy to conclude that the uncertainty on the cluster metallicity does not
affect significantly the evaluation of $L_{3.85}^{zahb}$. However, it is important to notice that
in all previous discussion we have used the spectroscopical measurement of $[Fe/H]$ as representative of
the {\sl true} cluster metallicity but, as it is well known (see, e.g. the review by Wheeler, 
Sneden \& Truran 1989), there are clear observational
evidence that the $\alpha-$elements are enhanced in GCs stars. 

So it is interesting to evaluate the
effect of the $L_{3.85}^{zahb}$ measurements due to an $\alpha-$elements enhancement. Due to the
lack of recent $\alpha-$element enhancement measurements for the selected clusters, we have adopted 
for all clusters (irrespective of their $[Fe/H]$ value) a mean enhancement $[\alpha/Fe]=0.30$ 
(Gratton et al. 1997), obtaining an average decreasing of $\log{L_{3.85}^{zahb}}$ of the order of 
$0.004-0.005$;
\smallskip\noindent
{\sl e)} as far as it concerns the RR Lyrae temperature scale, the accuracy of the $T_{eff}$ scale, adopted in
the present work, has been discussed in detail by DS96 and briefly reviewed in section III.
Therefore, now we do not repeat this analysis. However, there is an important question on the
temperature scale which we now wish to address concerning the possibility that our scale (as given by
the equation 4) can be dependent on the stellar luminosity. It is evident that this is a quite important
question for our investigation in fact if the adopted temperature scale should be dependent on the
luminosity of the RR Lyrae stars, this occurrence could produce a not accurate estimate of the ZAHB
luminosity.

On this point, Catelan (1998) and Catelan, Sweigart \& Borissova (1998, hereinafter CSB98) have 
recently claimed that \lq{\sl a relationship involving only the equilibrium temperature, blue 
amplitude and metallicity would be safer to adopt in period-shift analysis than CSJ92's 
{\rm (depending also on the pulsational period)} since period shifts caused by luminosity variations 
could easily be misinterpreted  as being due to temperature variations}\rq. 
So we have decided to check if the adopted temperature scale could be
dependent on the luminosity. For this aim, we have taken into account the sample of 17 field {\sl ab} type 
RR Lyrae, studied by using the Baade-Weesselink method by CSJ92, and investigated if the difference
between the temperature, provided by equation 4), and the one given by CSJ92 (their table 4) depends
on the measured luminosity of these stars. The result of such comparison is shown in figure 12)
\figure{12}{S}{80mm}{\bf Figure 12. \rm The residual between the effective temperature provided by
equation 4) and the temperature measured by CSJ92 as a function on the luminosity for the sample of
field RR Lyrae stars studied by CSJ92. The open circles represent the residuals corresponding to two
different solutions provided by the Baade-Wesselink analysis for the star SS Leo (see CSJ92 for more
details).}

It is evident from this plot the absence of any correlation between the effective temperature
residuals and the luminosity. Such occurrence clearly allows us to be confident in the use of DS96's
temperature scale in our analysis. However, to provide a more deep investigation on the
accuracy of our method, we have decided to repeat our investigation by using a different
temperature scale and, being aware of the warning from Catelan (1998),
the period independent CSB98 scale has been adopted.
\figure{13}{S}{80mm}{\bf Figure 13. \rm As figure 10) but for the cluster M3 but using the temperature scale
as given by CSB98.}

It is worth stressing that for all clusters in our sample, the estimates of the absolute bolometric 
magnitude of the ZAHB $\log{L_{3.85}^{zahb}}$ are in fine agreement with the one determined by using the
temperature scale provided by the equation 4). This occurrence provides a further plain evidence of the
reliability of both our temperature scale and our global approach.
The results of this investigation for the case corresponding to the cluster M3 (but the same results
have been achieved for the other clusters) have been plotted in figure 13). From this figure, one can
obtain the following indications:
\smallskip\noindent
i) the estimated  ZAHB luminosity level is equal to the one obtained by adopting the $T_{eff}$ given by
equation 4;
\smallskip\noindent
ii) the observational points, corresponding to the stars at the lower effective temperatures inside the
instability strip, as given by the CSB98 temperature scale are not in satisfactory agreement with the
pulsational theory prescriptions. In fact, it is worth noting that the slope of the observational data is
different from the theoretical one. In our belief, this occurrence has to be related to the fact
that the calibrating stars used by CSB98 for deriving their relation, do not cover the full expected
range of RR Lyrae effective temperatures.

\subsection{Comparison between field and cluster RR Lyrae stars}
\tx

In his accurate analysis on field RR Lyrae stars, whose parallaxes have been recently provided by the
Hipparcos mission, Catelan (1998) has clearly shown that it does not exist any evidence for a
mean luminosity difference between field RR Lyrae stars and GCs variables. This result has been
confirmed also by the work of Carney \& Lee (1998 - see the note added in proof in the Catelan's paper).
Since the main goal of the present investigation is to evaluate the ZAHB luminosity level from the
pulsational properties of RR Lyrae stars, it is natural to extend our investigation to field variables in
order to eventually provide further support to the results obtained by Catelan (1998). By courtesy of
Dr. M. Catelan, we have been provided with his original list of field RR Lyrae stars and related
pulsational properties, which corresponds to a sub-sample of the field variables, belonging to the 
original list of stars adopted by Tsujimoto et al. (1998) in their Hipparcos-based investigation. 
As far as it concerns the selection criteria adopted by Catelan (1998) to choose this sample of
variables, we refer the interested reader to the quoted paper.
\figure{14}{S}{100mm}{\bf Figure 14. \rm Panel a) Comparison in the period - effective temperature
diagram between {\sl ab} RR Lyrae stars in the cluster M3 and field variables whose parallaxes have been
obtained with the Hipparcos satellite (the metallicity range covered by such stars is labelled). Panel
b) the same of panel a) but for variables in the clusters M15 and M68 and field variables in a different
(as labelled) metallicity range.}

The procedure adopted to perform the comparison between GCs RR Lyrae stars and field variables is quite
similar to the one used by Catelan (1998): we have split the data for field variables in two different
subsamples corresponding to two different metallicity ranges, with an average metallicity equal to 
$\approx-2.0$ and $\approx-1.3$, respectively. Then we have compared the pulsational properties of field
variables in each range of metallicity with the properties of RR Lyrae stars in GC of suitable
metallicity. In figure 14a), it is shown the comparison between variable in the cluster M3 and field RR
Lyrae stars filling in the \lq{\sl metal-rich}\rq\ range; while figure 14b) shows the same comparison
but between RR Lyrae stars in M15 and M68 and field stars in the \lq{\sl metal-poor}\rq\ range.

From the results shown in these figures, it is evident that there is no clear difference in the
pulsational properties of GCs RR Lyrae stars and field ones in both (at least) explored metallicity range,
which could be related to the existence of a {\sl real} difference in luminosity between GCs and field
variables. This result clearly provides further support to the Catelan's (1998) investigation.
\figure{15}{S}{110mm}{\bf Figure 15. \rm Panel a) Comparison between the estimates of the parameter 
$\log{L_{3.85}^{zahb}}$ for different GCs, as provided by the present analysis by assuming a scaled solar
heavy elements distribution and theoretical expectations on this quantity as given by different authors 
(see labels and the text for more details). Panel b) as panel a) but by adopting for the selected clusters an
average $\alpha-$elements enhancement $\rm [\alpha/Fe]=0.30$.}

\subsection{Comparison with theoretical results.}
\tx

Since we have been able to evaluate the bolometric magnitude of the ZAHB for different clusters, it
is now obvious to compare present results with the theoretical prescriptions on this important
quantity. This comparison has been performed in figure 15), where we have plotted the evaluations on 
$\log{L_{3.85}^{zahb}}$ as obtained in the present work and also the most recent evolutionary 
theoretical evidences. In particular, we have considered the ZAHB stellar models from 
Castellani et al. (1991), Dorman, Rood \& O'Connell (1993), Caloi et al. (1997), 
Cassisi \& Salaris (1997), Cassisi et al. (1998a,b), Salaris \& Weiss (1998) and Vandenberg (1998). 
All these models have been computed in a canonical evolutionary framework,
but the Cassisi et al.'s (1998a,b) ones which have been computed for both a canonical scenario and
also a scenario accounting for element (Helium + heavy elements) diffusion. We refer to the quoted
papers for details on the evolutionary computations performed by the various authors. 
It is worth noticing that the ZAHB models of Cassisi \& Salaris (1997) are full consistent with the
evolutionary computations used in the previous sections.

We have already discussed the effect on the evaluation of the ZAHB luminosity due to a possible
$\alpha-$elements enhancement in the heavy elements distribution and shown that it is very little. 
Nevertheless, when comparing observational evidences with theoretical results, one has to pay 
attention to adopt self-consistent metallicity evaluations. For such reason and also being aware for the
lack of accurate $\alpha-$elements enhancement measurements (consistent with the [Fe/H] metallicity scale
of Carretta \& Gratton 1997) for the clusters in our sample, we have made two different assumptions by 
adopting both $\rm [\alpha/Fe]=0.0$ and 0.30 (following the suggestion given by Gratton et al. 1997).
The comparison between theory and present results for both assumptions on the $\alpha-$elements enhancement,
has been performed in figures 15a) and 15b), respectively.

From this comparison, in the limits of the little sample of selected clusters, it is possible 
to derive the following indications:
\medskip\noindent
{\sl i)} the slope of $\log{L_{3.85}^{zahb}}$ with the metallicity is in satisfactory
agreement, at least for $[M/H]\le-1.0$, with the values provided by almost all theoretical investigations,
but the models provided by Caloi et al. (1997);
\smallskip\noindent
{\sl ii)} in the more metal-rich range, it seems to exist a significant discrepancy between theory and the
\lq{\sl pulsational}\rq\ estimations if the observational point corresponding to the cluster NGC6171 is
accounted for. The discrepancy appears less evident when considering an $\alpha-$elements enhancement for the GCs
equal to 0.30. Nevertheless, it is worth noticing that we have only two
clusters: NGC6171 and NGC6362 at higher metallicity and, unfortunately the evaluation of $\log{L_{3.85}^{zahb}}$ for 
NGC6171 is affected by a larger uncertainty. 
However, if further investigations should support this preliminary results, one is
facing on with the evidence that the theoretical models underestimate, in comparison with the observations, 
the decrease of the ZAHB luminosity - at a fixed effective temperature - when the metallicity increases. 
Nevertheless, when excluding from the analysis
this largely deviant result, it is still possible to assess that theory and observations are in fair agreement at least
up to a metallicity of the order of $\approx-0.9, -0.8$ dex;
\smallskip\noindent
{\sl c)} it is worth noting the satisfactory agreement which has been achieved for both assumptions on the
$\alpha-$elements enhancement between the values of 
$\log{L_{3.85}^{zahb}}$ for the selected clusters and the evolutionary prescriptions as given by Cassisi \&
Salaris (1997) and Vandenberg (1998) concerning both the slope and the absolute values;
\smallskip\noindent
{\sl d)} recently Cassisi et al. (1998a,b), in order to thoroughly investigate the effects on several
important evolutionary quantities of the most updated evaluations on the several physical inputs adopted in
stellar computations, have provided a complete set of evolutionary models for both H and He burning phase
in a canonical scenario and also by accounting for element diffusion. The results shown in figure
15a), seem to indicate that the ZAHB luminosity levels provided by these models appear to be ruled out by the
present investigations. The situation is slightly better when considering models accounting for element
diffusion and when assuming for the clusters an $\rm [\alpha/Fe]$ value larger than zero.
Such occurrence could be regarded as an indication of the fact that one (or more than one) updated 
physical inputs (as for instance, nuclear cross sections, neutrino energy losses, conductive opacity and so
on) adopted in these computations, is {\sl still} affected by a large uncertainty. 

Nevertheless, we think that, due to the large uncertainties on the observational measurements of both
[Fe/H] and the heavy elements distribution, there are not yet clear evidences that this is the case.
Even if, a deeper insight on this topic is clearly out of the aim of the present work, we wish only to
notice, as a warning, that it has been recently discussed (Fiorentini, Lissia \& Ricci 1998) the possibility that 
the element diffusion coefficients adopted in the evolutionary codes, could be affected by a large uncertainty 
and it has been also shown by Castellani \& Degl'Innocenti (1998) that these uncertainties can produce significant
changes on the ZAHB luminosity. Therefore, we think that it does {\sl still} exist in the parameters 
(adopted in stellar computations) space the possibility to obtain a better agreement between present
results and these evolutionary computations. 

As a final point, we wish to notice that the present results allow us to obtain relevant informations
on both the slope and the zero point of the relation between the absolute visual magnitude of the RR Lyrae
stars and the metallicity on a ground quite independent from similar analysis already performed on this
concern. However, a deep insight on this topis is out of the aim of the present work so we will address
it in a forthcoming paper (Cassisi \& De Santis 1998).

It is evident that safer conclusions about the real luminosity level of the ZAHB could be obtained only 
when the sample of GCs with high accuracy photometry for both RR Lyrae stars and non-variable HB
structures and a fine measurement of the RR Lyrae pulsational properties, will be larger. Since in the
present work it has been clearly shown that the analysis of the pulsational properties of RR Lyrae stars
allows us to evaluate the ZAHB luminosity with high accuracy and with an approach largely independent from
evolutionary computations, this occurrence is strongly desired in order to finally achieve a large consensus
on the most important Population II distance scale.

\section*{Acknowledgments}
\tx 

The authors are deeply grateful to Giuseppe Bono for a detailed reading of this manuscript as well as
for several suggestions and interesting discussions on this topic all along these years.
We warmly thanks Marcio Catelan for kindly providing the data on field RR Lyrae stars adopted in the present
work and, also for interesting discussion on this topic. S.C. wish also to acknowledge Don Vandenberg for 
sending his updated stellar models in advance on publication as well as for sharing useful informations on
evolutionary computations. We thank Alistair Walker for kindly supplying his data for the RR Lyrae stars in 
NGC6362 in advance on publication. It is a pleasure to warmly thank Maurizio Salaris for an accurate
reading of a preliminary draft as well as for all interesting discussions on this topic. We would like to thank
our referee for the pertinence of her/his suggestions.

\section*{References}
\bibitem Bono, G., Caputo, F., Castellani, V. \& Marconi, M. 1997, A\&AS, 121, 327
\bibitem Bono, G., Caputo, F. \& Stellingwerf, R.F. 1994, AJ, 423, 294
\bibitem Bono, G., Caputo, F. \& Stellingwerf, R.F. 1995, ApJSS, 99, 263
\bibitem Bono, G., Cassisi, S. \& Castellani, V. 1998, {\sl in preparation}
\bibitem Bono, G. \& Stellingwerf, R.F. 1994, ApJS93, 233
\bibitem Brocato, E., Castellani, V., Raimondo, G. \& Walker, A.R. 1998, {\sl in preparation} 
\bibitem Buser, R. \& Kurucz, R.L. 1978, A\&A, 70, 555 (BK78)
\bibitem Caloi, V., D'Antona, F. \& Mazzitelli, I. 1997, A\&A, 320, 823
\bibitem Caputo, F., Castellani, V., Marconi, M. \& Ripepi, V. 1998b, MNRAS, {\sl submitted to}
\bibitem Caputo, F. \& De Santis, R. 1992, AJ, 104, 253 (CDS92)
\bibitem Caputo, F., Marconi, M. \& Santolamazza, P. 1998a, MNRAS, 293, 364
\bibitem Carney, B.W., Storm,J. \& Jones, R.V. 1992, ApJ, 386, 663 (CSJ92)
\bibitem Carretta, E. \& Gratton, R.G. 1997, A\&AS, 121, 95 (CG97)
\bibitem Cassisi, S., Castellani, V., Degl'Innocenti, S. \& Weiss, A. 1998a, A\&AS, 129, 267
\bibitem Cassisi, S., Castellani, V., Degl'Innocenti, S., Salaris, M. \& Weiss, A. 1998b, A\&AS, {\sl in press}
\bibitem Cassisi, S. \& De Santis 1998, {\sl in preparation}
\bibitem Cassisi S. \& Salaris M. 1997, MNRAS, 285, 593
\bibitem Castellani, V., Chieffi, A. \& Pulone, L. 1991, ApJS, 76, 911 
\bibitem Castellani, V. \& Degl'Innocenti, S. 1998, private communication
\bibitem Castellani, V. \& De Santis, R. 1994, ApJ, 430, 624
\bibitem Castelli, F., Gratton, R.G. \& Kurucz, R.L. 1997a, A\&A, 318, 841 (CGK97)
\bibitem Castelli, F., Gratton, R.G. \& Kurucz, R.L. 1997b, A\&A, 324, 432 (CGK97)
\bibitem Catelan, M. 1998, ApJ, 495, L81
\bibitem Catelan, M., Sweigart, A.V. \& Borissova, J. 1998, in ASP Conf. Ser. 135, {\sl A half century
of stellar pulsation interpretations: a tribute to Arthur N. Cox}, ed. P.A. Bradley \& A. Guzik, (San
Francisco: ASP), p. 41 (CSB98)
\bibitem Chaboyer B., Demarque P., Kernan P.J. \& Krauss L.M. 1998, ApJ, 494, 96
\bibitem De Boer, K.S., Tucholke, H.-J. \& Schmidt, J.H.K. 1997, A\&A, 317, L23
\bibitem De Santis, R. 1996, A\&A, 306, 755 (DS96)
\bibitem Dorman, B., Rood, R.T. \& O'Connell, R.W. 1993, ApJ, 419, 596
\bibitem Feast M.W. \& Catchpole R.M. 1997, MNRAS, 286, L1
\bibitem Fernley, J., Barnes, T.G., Skillen, I., Hawley, S.L., Hanley, C.J., Evans, D.W., Solano, E. 
\& Garrido, R. 1998, A\&A, 330, 515
\bibitem Fiorentini, G., Lissia, M. \& Ricci, B. 1998, A\&A, {\sl submitted to}, Astro-ph/9808010
\bibitem Gould, A. \& Uza, O. 1998, ApJ, 494, 118
\bibitem Gratton, R.G. 1998, MNRAS, 296, 739  
\bibitem Gratton, R.G., Fusi Pecci, F., Carretta, E., Clementini, G.,
Corsi, C.E. \& Lattanzi, M. 1997, ApJ, 491, 749
\bibitem Green E.M. 1988, in "Calibration of Stellar Ages", A.G. Davis Philip ed. 
(L. Davis Press) p. 81 (Yale)
\bibitem Kovacs, G., Buchler, J.R. \& Marom, A. 1991, A\&A, 252, L27
\bibitem Kurucz R.L. 1979, ApJS, 40, 1
\bibitem Kurucz, R.L. 1992, in IAU Symp. 149, \lq{The stellar populations of galaxies}\rq, Barbuy, B. \&
Renzini, A. eds., Dordrecht: Kluwer, p. 225 (K92)
\bibitem Lub, J. 1977, A\&AS, 29, 345
\bibitem Madore, B.F. \& Freedman, W.L. 1997, ApJ, 492, 110
\bibitem McNamara, D.H. 1997, PASP, 109, 1221
\bibitem Panagia, N. 1998, in IAU Symp. n. 190, \lq{New Views of the Magellanic Clouds}\rq,
p. 53
\bibitem Pont, F., Mayor, M., Turon, C. \& Vandenberg, D.A. 1998, A\&A, 329, 87
\bibitem Reid I.N. 1997, AJ, 114, 161
\bibitem Reid I.N. 1998, AJ, 115, 204
\bibitem Renzini, A. 1991, NATO ASI Series, in \lq{Observational Tests of Cosmological Inflation}\rq,
Shanks, T., Banday, A.J., Ellis, R.S., Frenk, C.S. \& Wolfendale, A.W. eds., Vol. 348, p. 131
\bibitem Rogers, F.J. \& Iglesias, C.A. 1992, ApJS, 79, 507
\bibitem Salaris, M. \& Cassisi, S. 1996, A\&A, 305, 858
\bibitem Salaris, M. \& Cassisi, S. 1998, MNRAS, 298, 166
\bibitem Salaris, M. \& Weiss, A. 1998, A\&A, 335, 943
\bibitem Sandage, A. 1981, ApJ, 248, 161
\bibitem Sandage, A. 1990, ApJ, 350, 631
\bibitem Sandage, A., Katem, N. \& Sandage, M. 1981, ApJS, 46 41
\bibitem Silbermann, N.A. \& Smith, H.A. 1995, AJ, 110, 704
\bibitem Tsujimoto, T., Miyamoto, M. \& Yoshii, Y. 1998, ApJ, 492, L79
\bibitem van Albada, T.S. \& Baker, N. 1971, ApJ, 169, 311
\bibitem Vandenberg, D.A. 1998, private communication
\bibitem Vandenberg D.A. \& Bell R.A. 1985, ApJS, 58, 561
\bibitem van Leeuwen, F., Feast, M.W., Whitelock, P.A. \& Yudin, B. 1997, MNRAS, 287, 955
\bibitem Walker, A.R. 1994, AJ, 108, 555
\bibitem Walker, A.R. 1998, AJ, 116, 220
\bibitem Wheeler J.C., Sneden C. \& Truran J. W. 1989, ARAA, 252, 179
\bibitem Zinn, R. \& West, M.J. 1984, ApJS, 55, 45
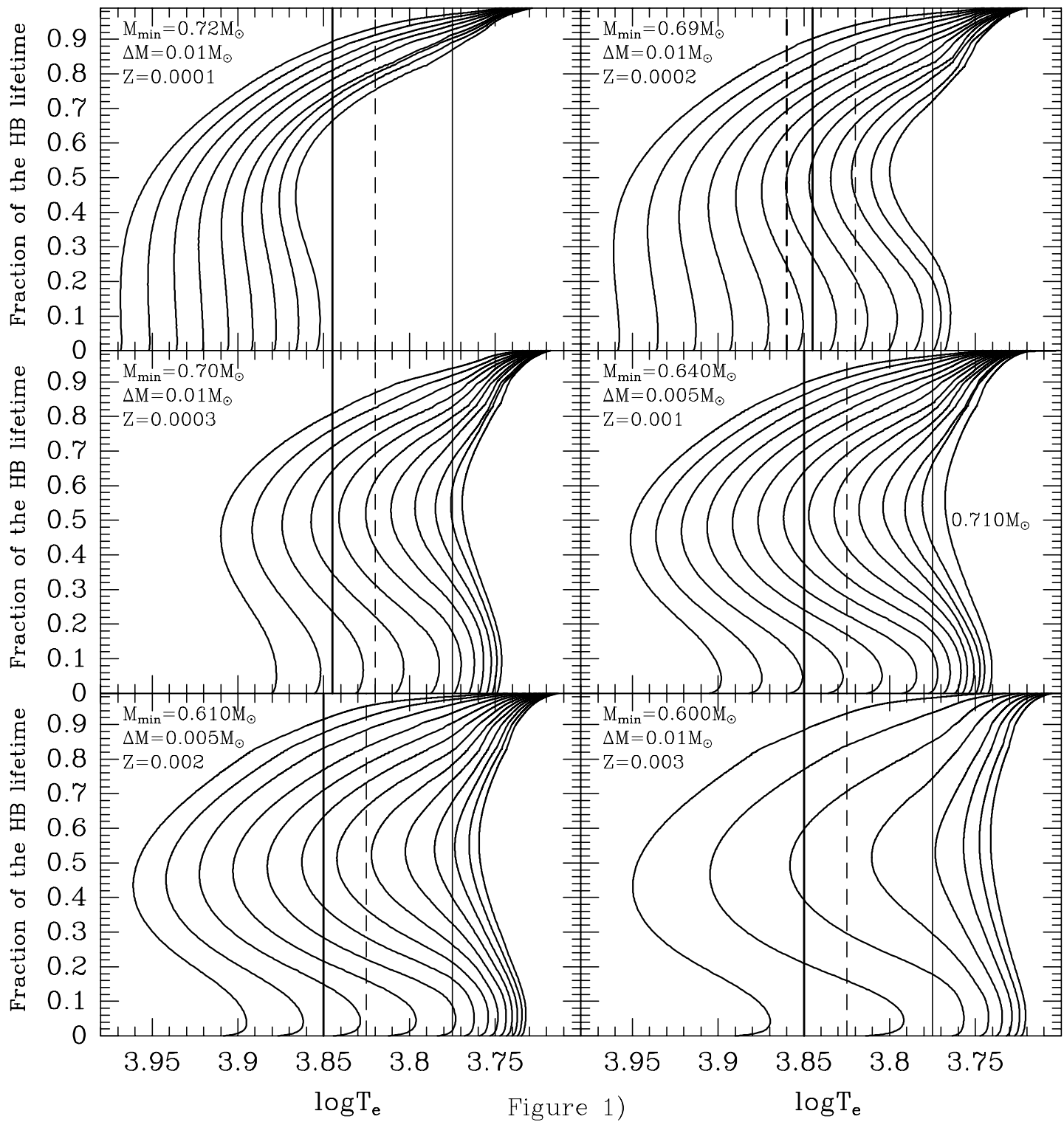
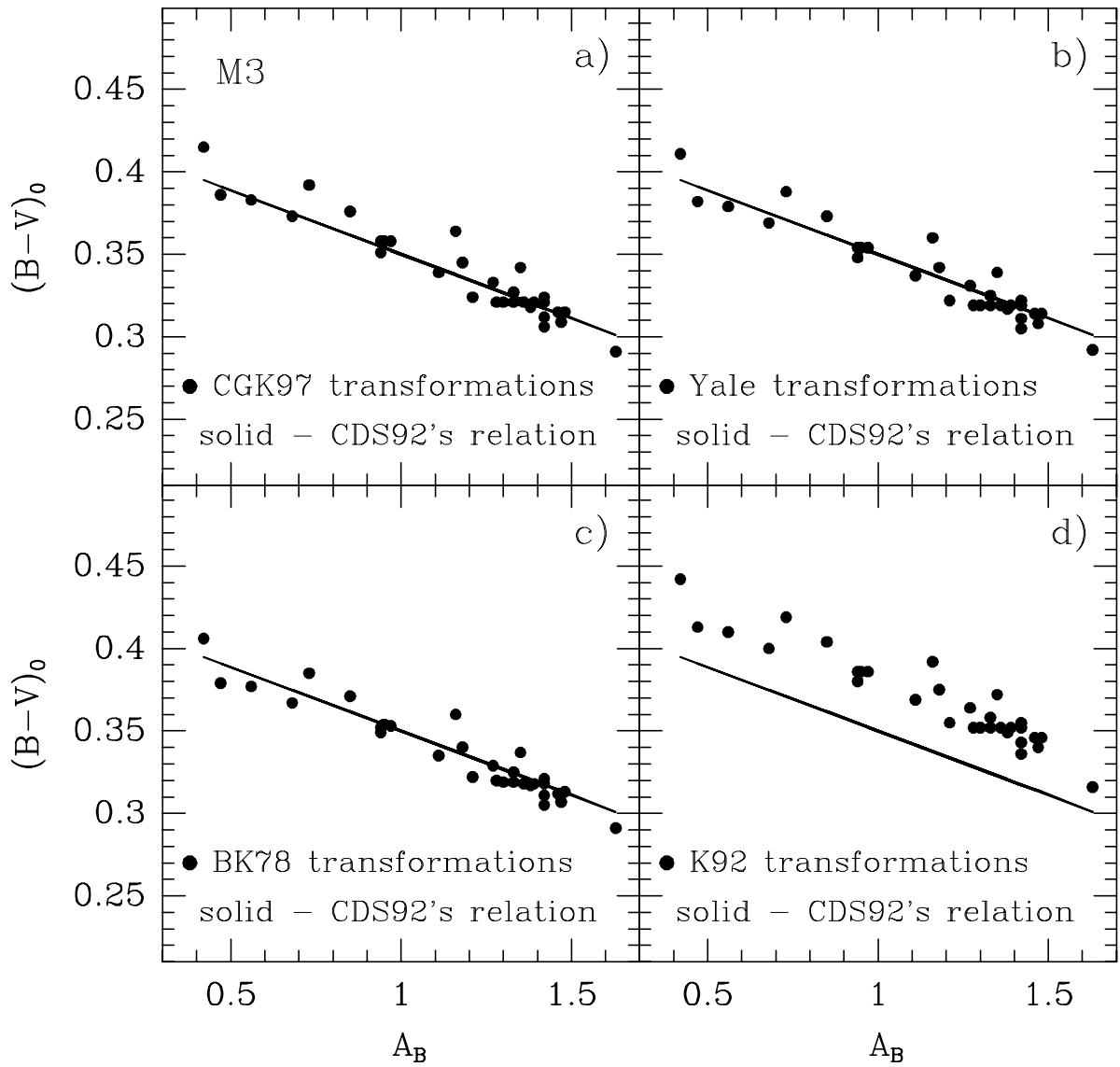
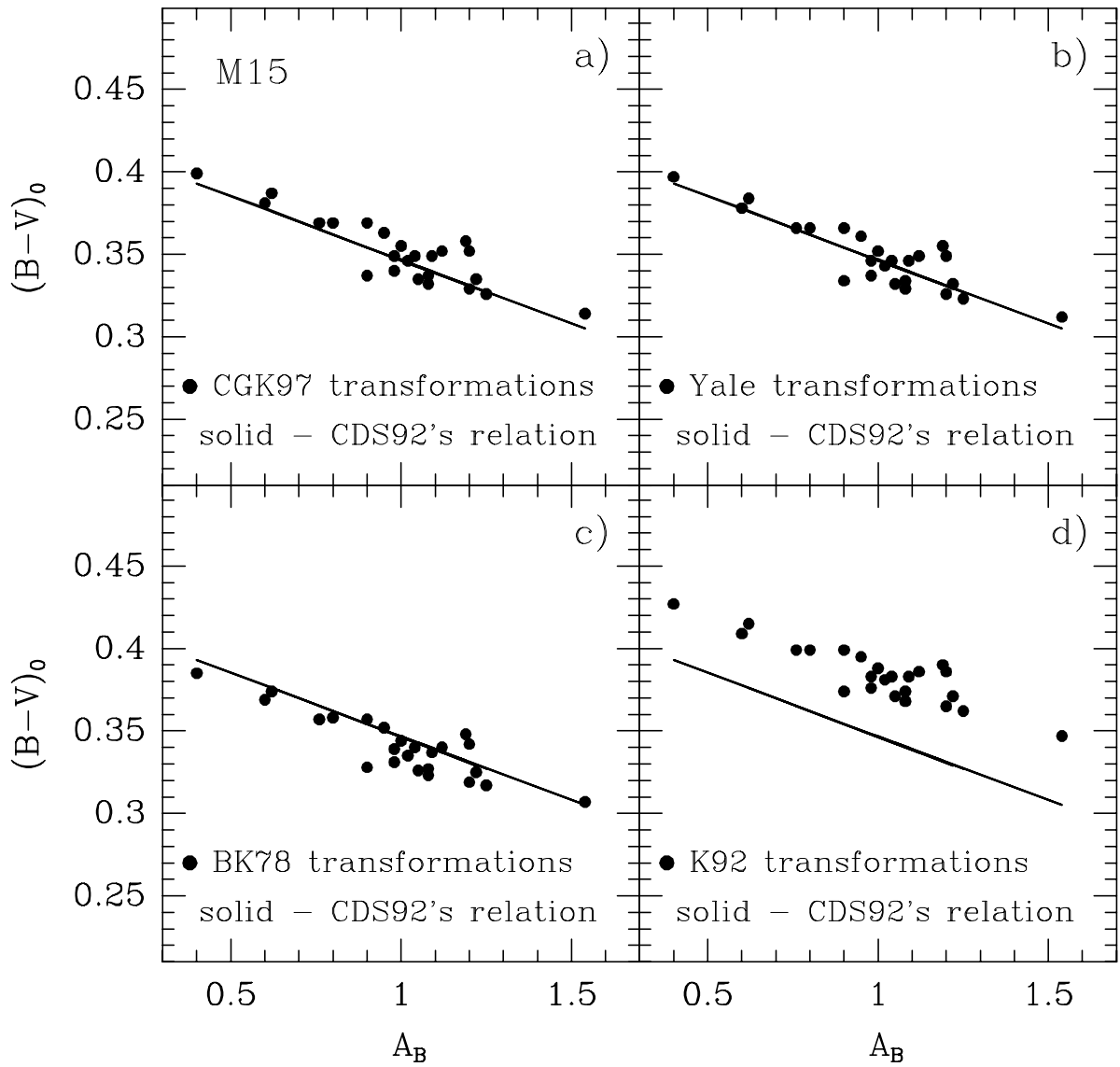
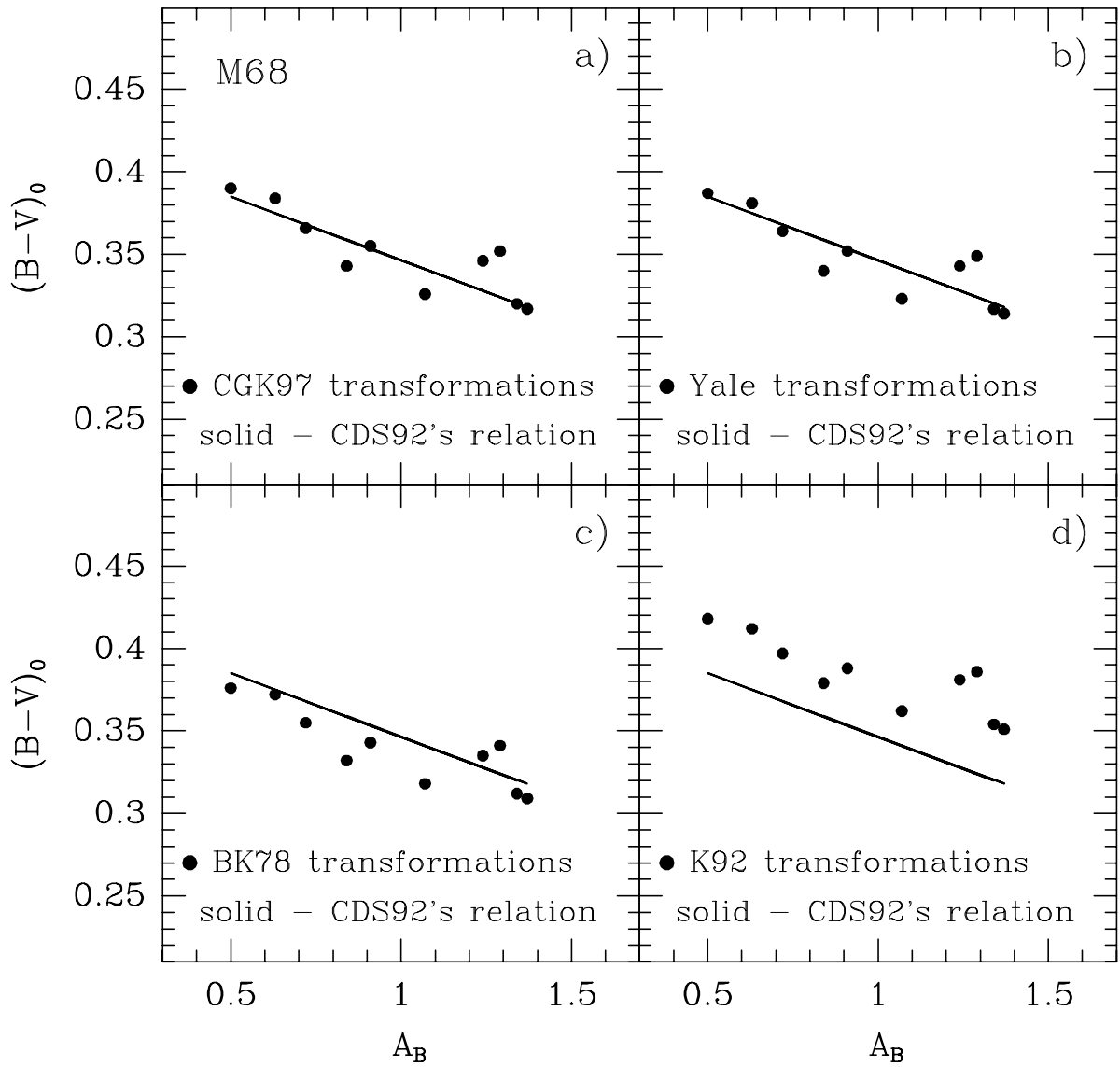
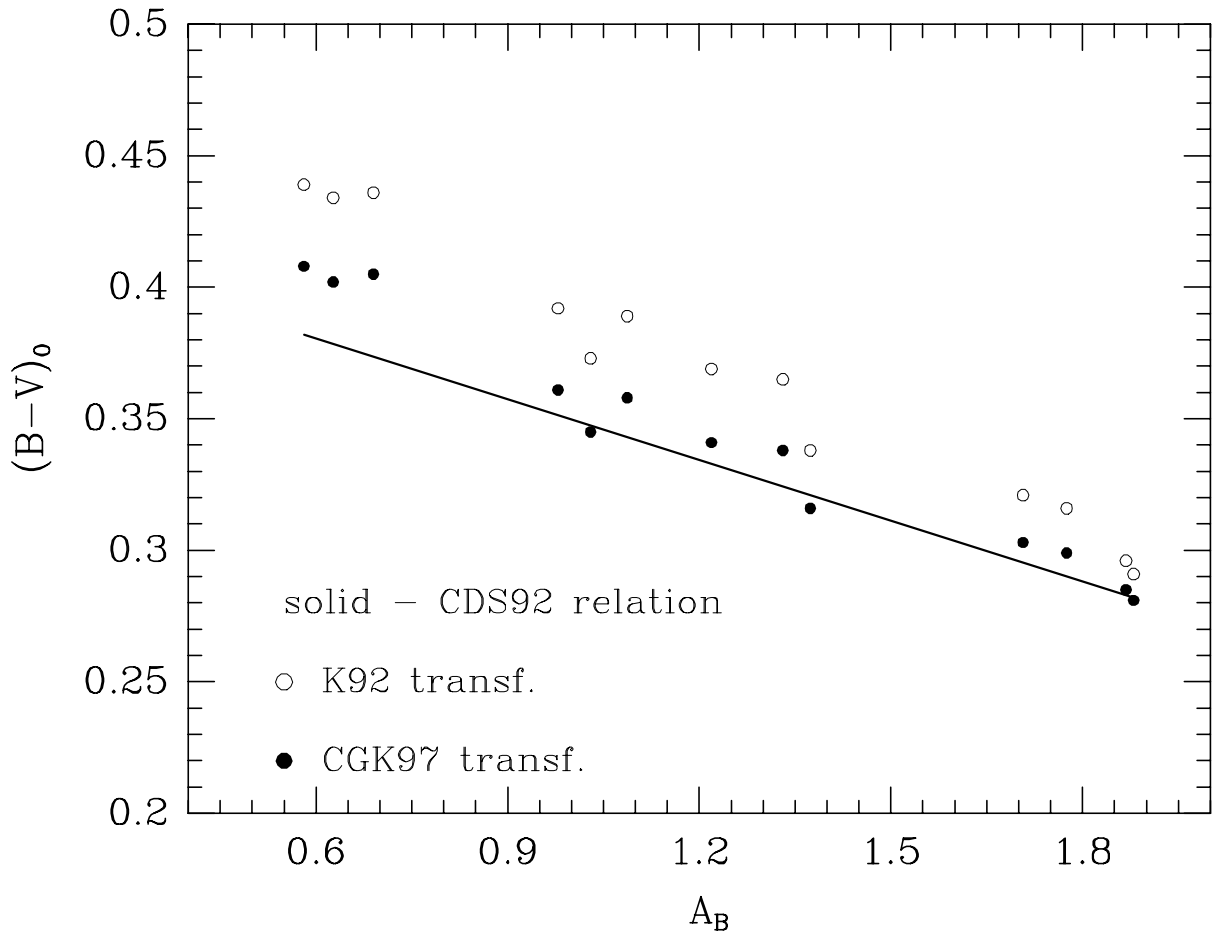
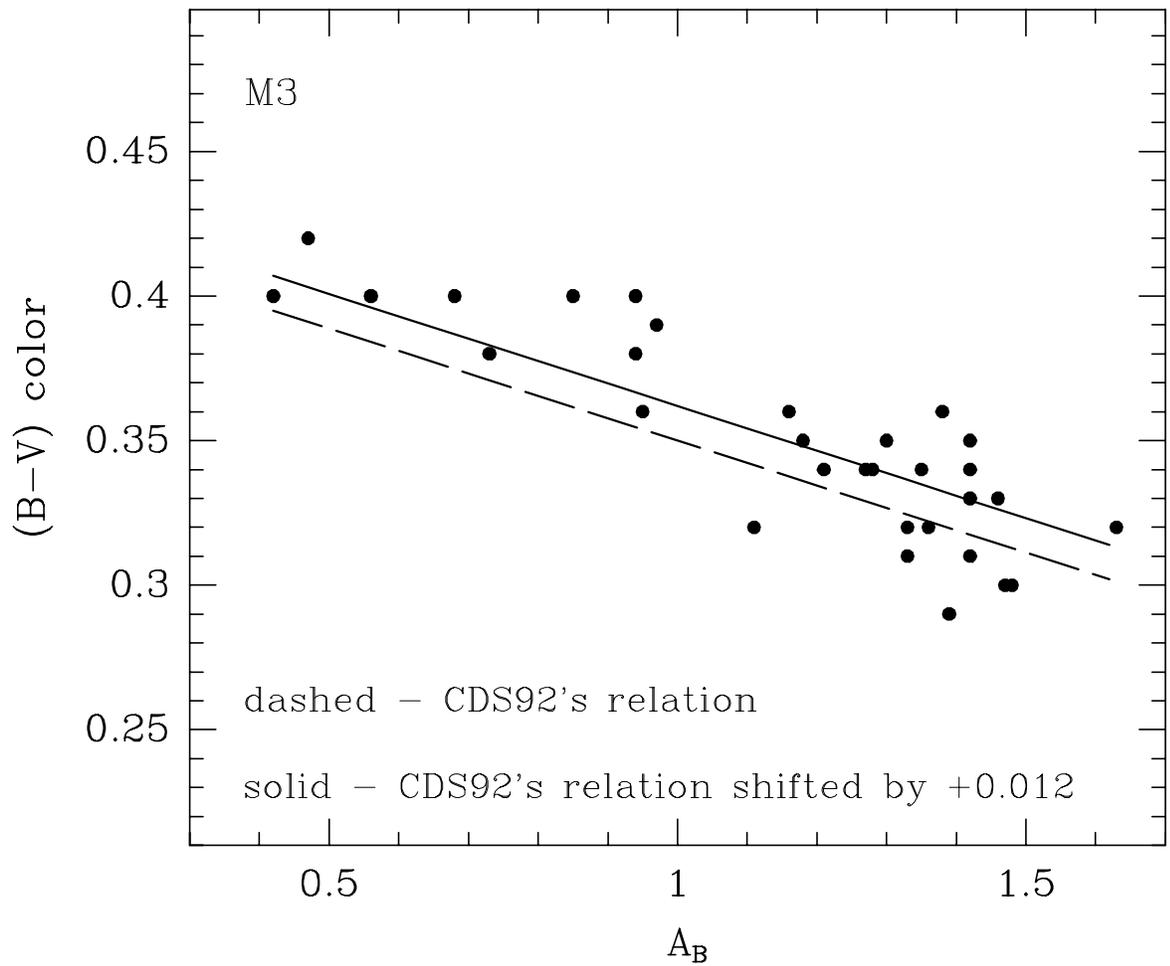
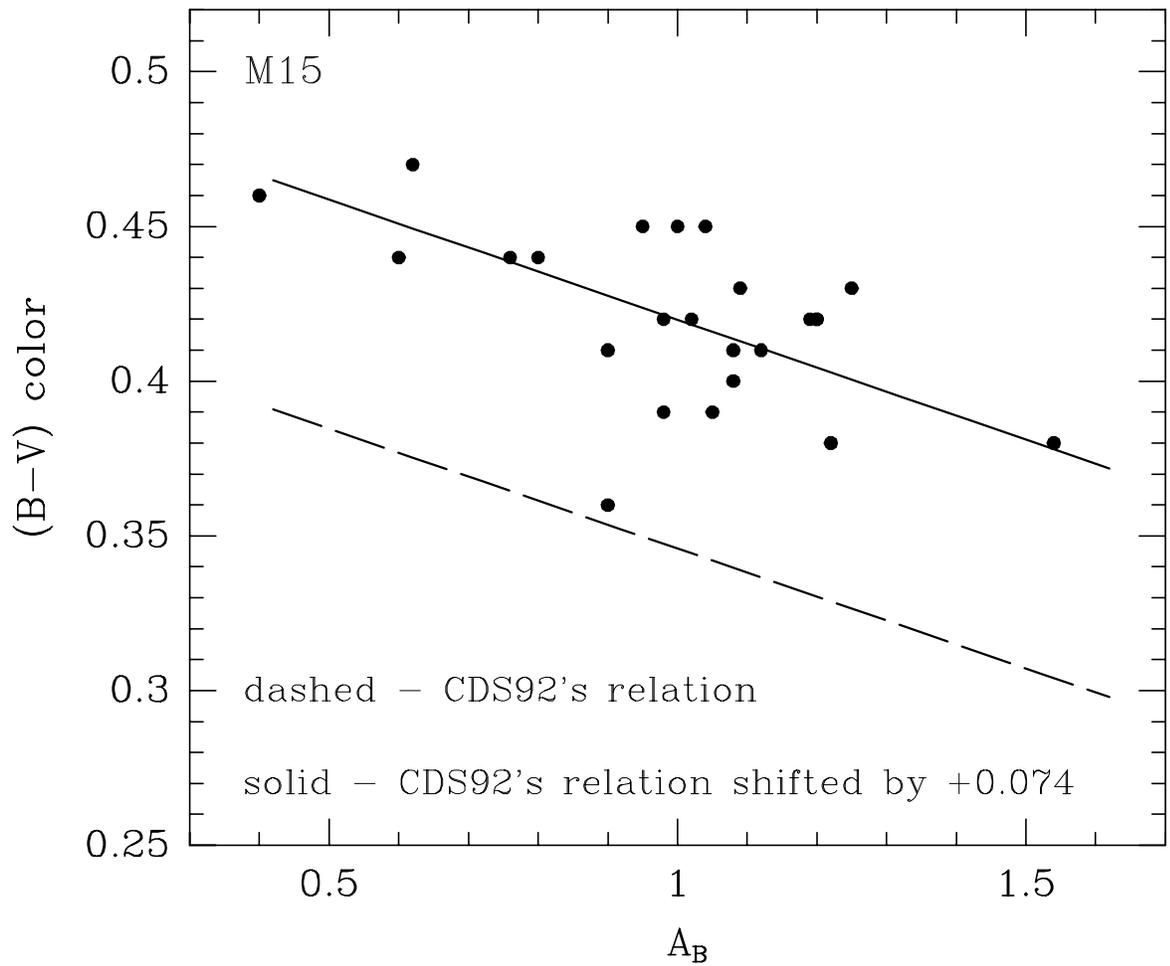
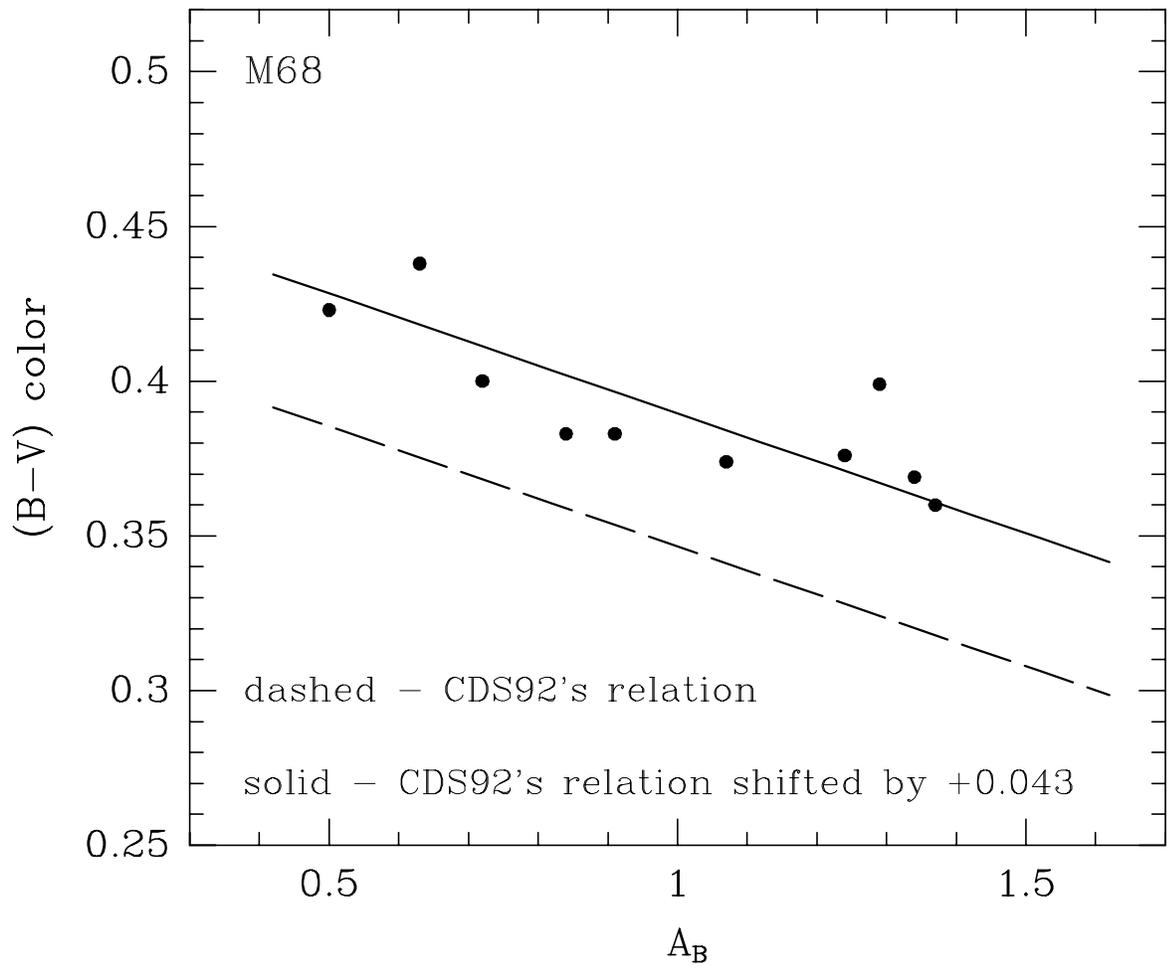
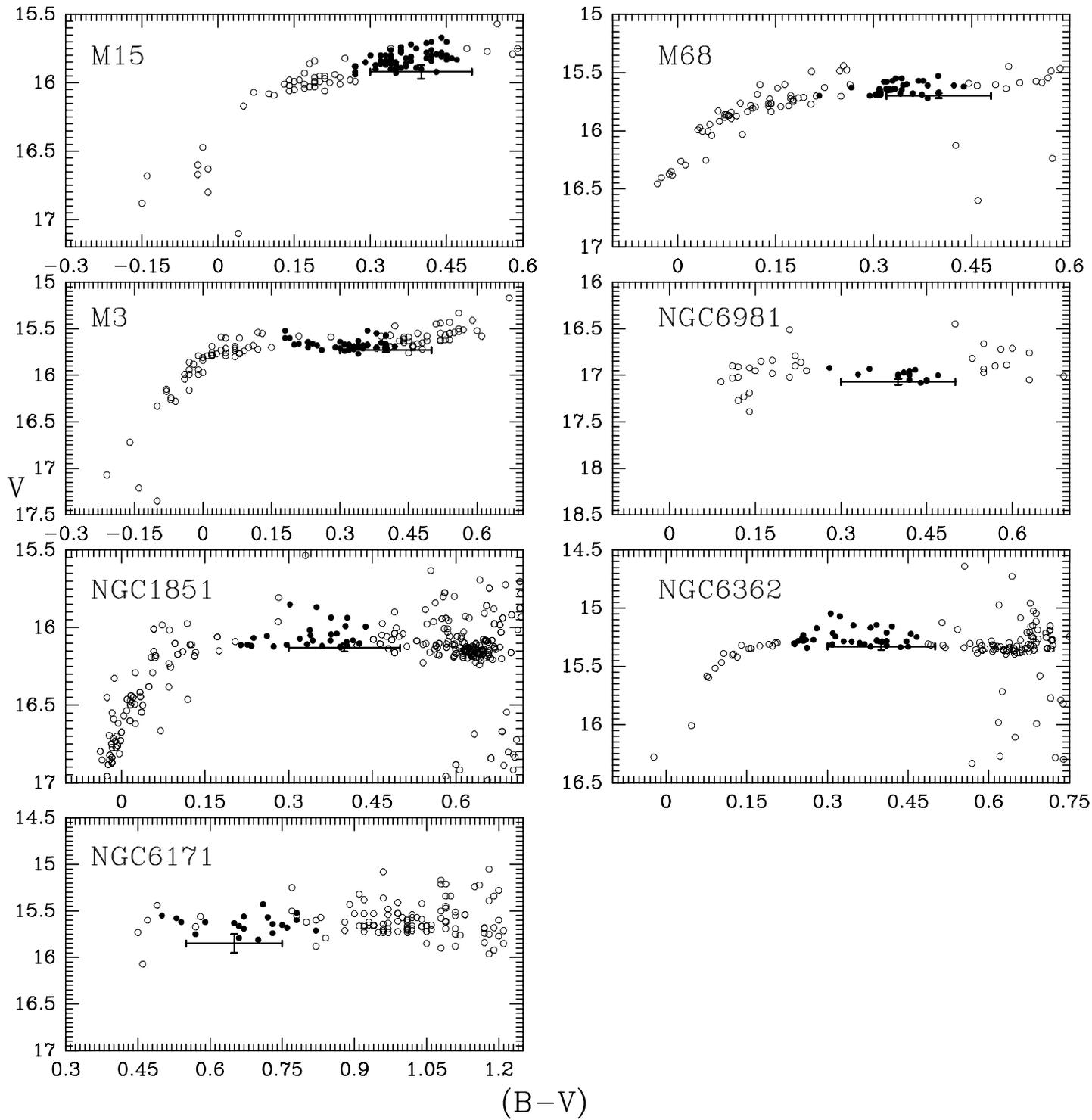
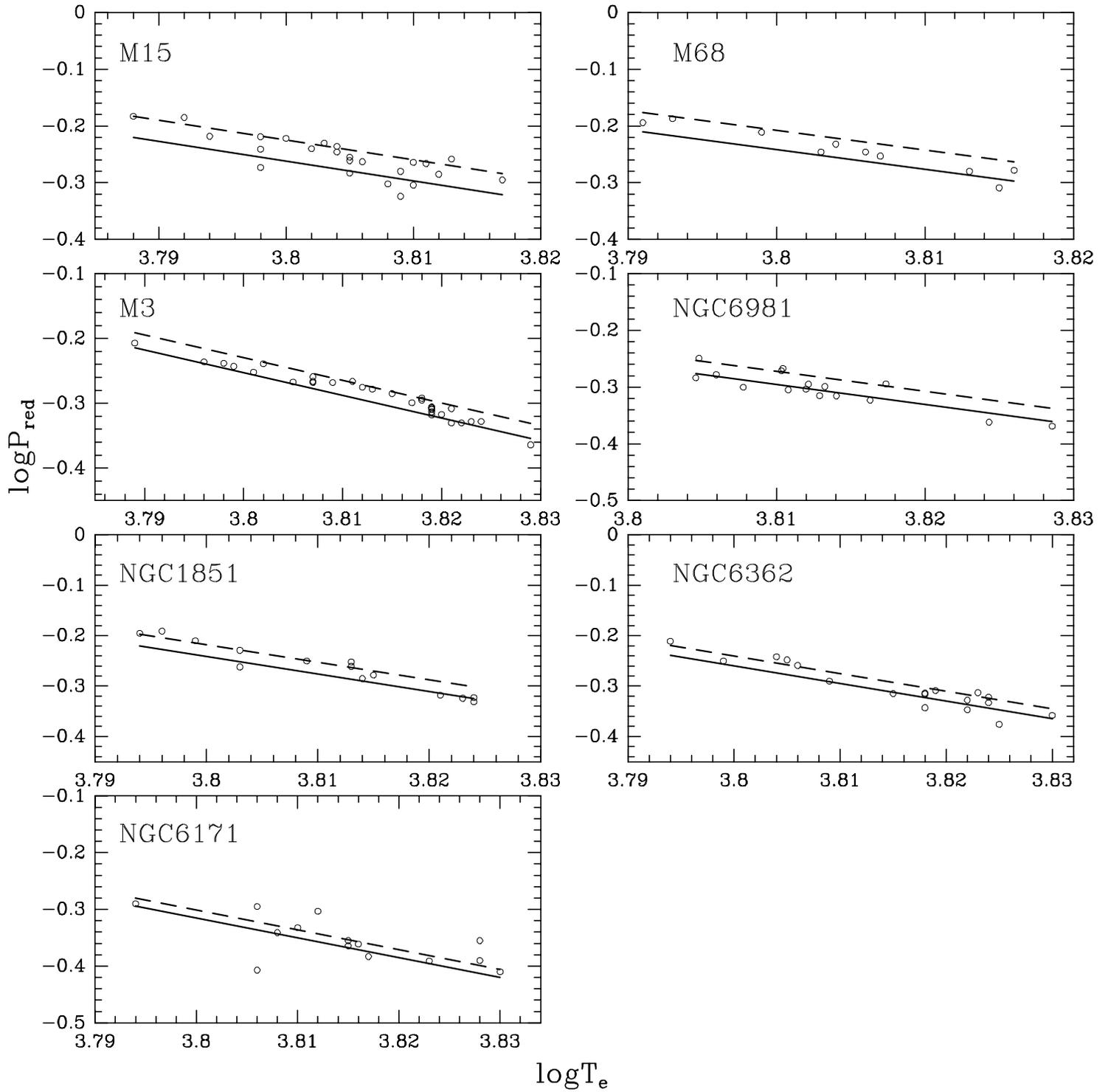
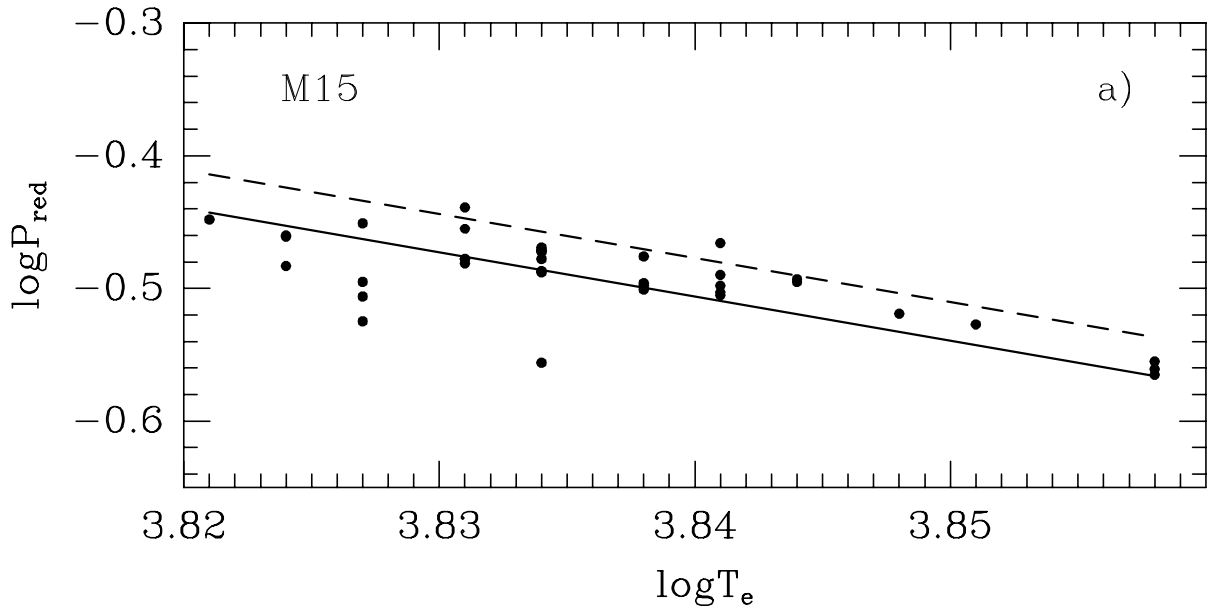
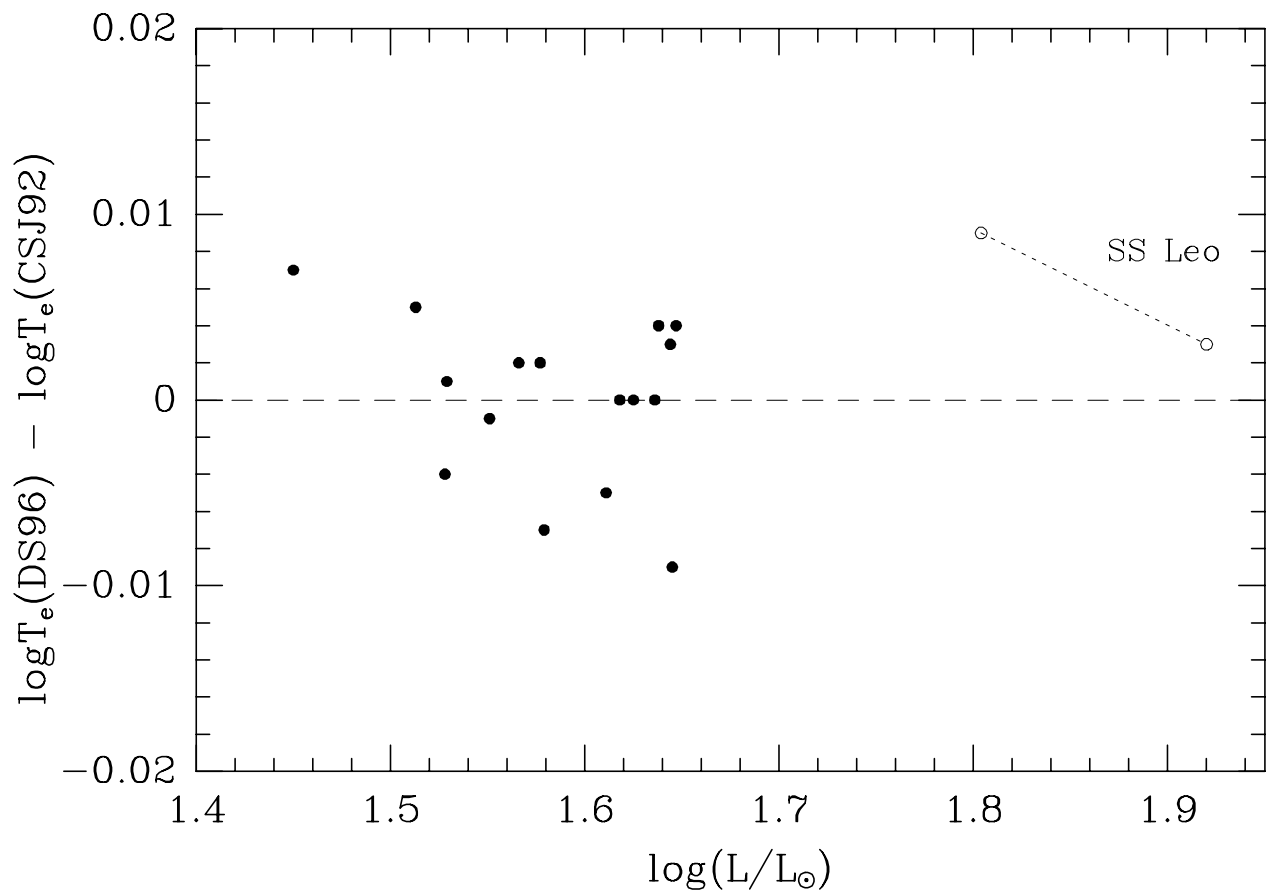
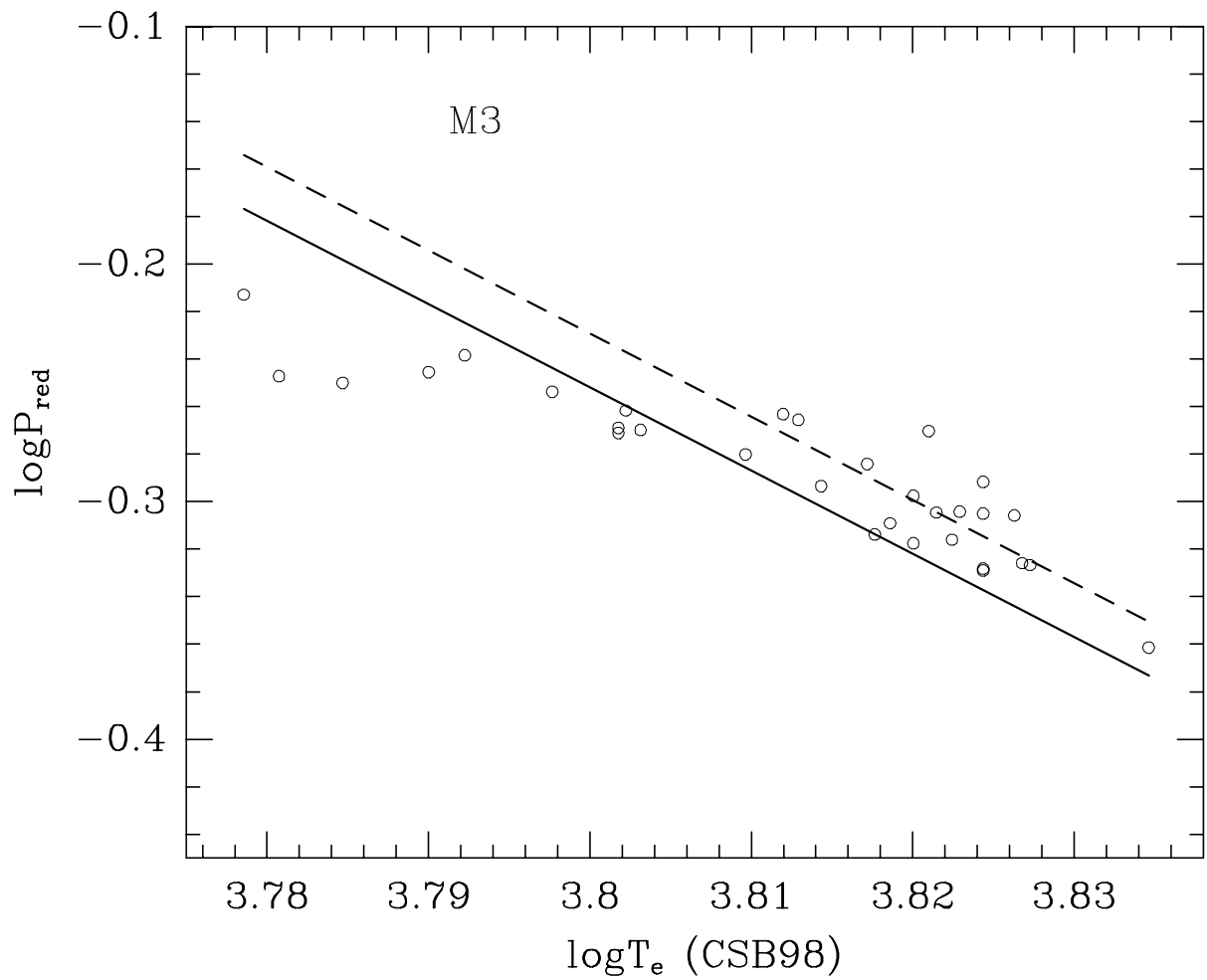
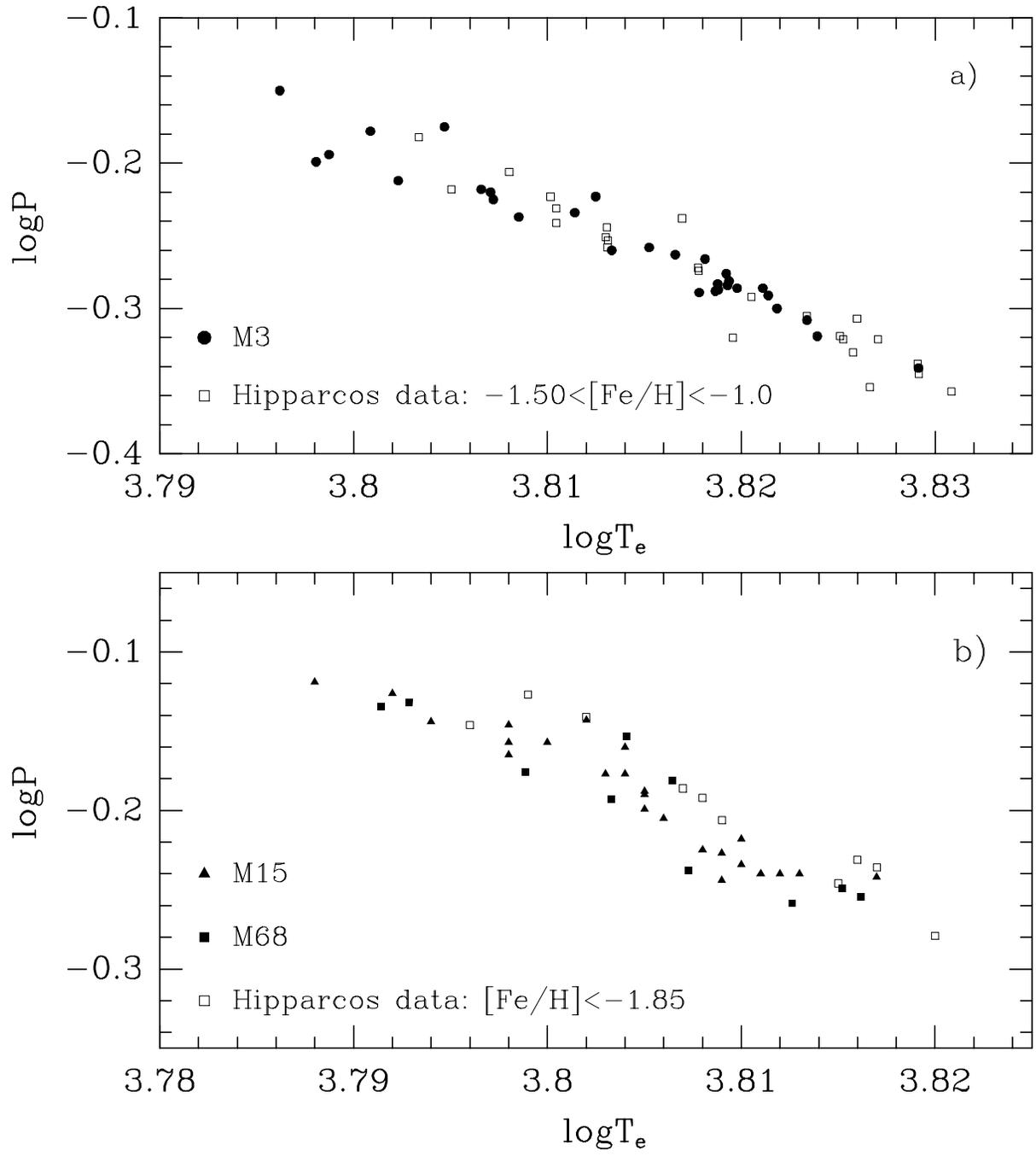
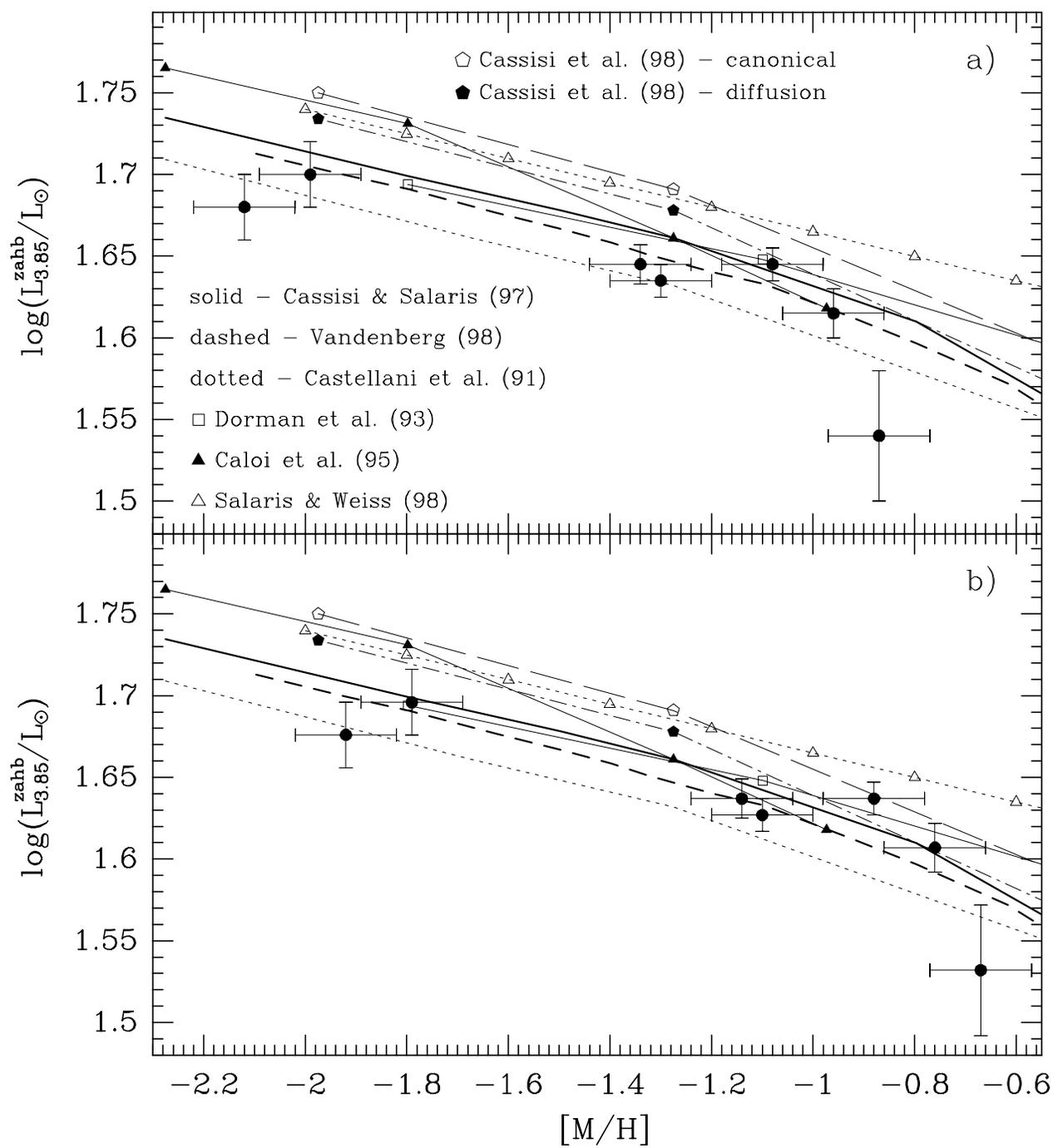
\bye